\documentclass{article}
\usepackage{arxiv}
\usepackage{amssymb}
\usepackage[figuresright]{rotating}

\usepackage{amsmath}
\usepackage{subcaption}
\usepackage{graphicx}
\usepackage{caption}
\usepackage{hyperref}
\usepackage{xcolor}
\usepackage{enumitem}

\setitemize{noitemsep,topsep=0pt,parsep=0pt,partopsep=0pt}

\newcommand{\FIGSCALE}{0.2}
\newcommand{\x}{\mathbf{x}}
\newcommand{\uu}{\mathbf{u}}

\title{The Use of the Simplex Architecture to Enhance Safety in Deep-Learning-Powered Autonomous Systems}
\author{
\makebox[\textwidth][c]{%
Federico Nesti\textsuperscript{1}\thanks{Corresponding author: federico.nesti@santannapisa.it} \quad
Niko Salamini\textsuperscript{1} \quad
Mauro Marinoni\textsuperscript{1}}\\[0.5ex]
\makebox[\textwidth][c]{%
\textbf{Giorgiomaria Cicero}\textsuperscript{1,2} \quad
\textbf{Gabriele Serra}\textsuperscript{2} \quad
\textbf{Alessandro Biondi}\textsuperscript{1,2} \quad
\textbf{Giorgio Buttazzo}\textsuperscript{1}}\\[1ex]
\textsuperscript{1}Department of Excellence in Robotics \& AI, Scuola Superiore Sant'Anna, Pisa, Italy \\
\textsuperscript{2}Accelerat S.r.l., Via L. Alamanni Lotto D/2, 5A, San Giuliano Terme, Italy
}

\begin{document}
\maketitle
%\begin{frontmatter}




\begin{abstract}
%[VIETATO USARE ACRONIMI!]
% 

%Abstract: 
%-	Current trustworthiness issues in deep learning
%o	Safety
%o	Security
%o	Predictability
%o	Explainability? (optional)
%-	solution: extension of the simplex architecture / runtime monitor to an architectural level
%o	Use of the hypervisor to isolate the non-trustworthy components 
%o	Use of a “certified” safety controller and safety monitor
%-	Contributions: application of the architecture in 2 use-cases:
%o	Pendulum
%o	Rover

%\TODO{I nomi delle componenti vanno decisi ed uniformati nel paper (rich os, safety monitor, safe controller and so on)}

Recently, the outstanding performance reached by neural networks in many tasks has led to their deployment in autonomous systems, such as robots and vehicles. However, neural networks are not yet trustworthy, being prone to different types of misbehavior, such as anomalous samples, distribution shifts, adversarial attacks, and other threats.
Furthermore, frameworks for accelerating the inference of neural networks typically run on rich operating systems that are less predictable in terms of timing behavior and present larger surfaces for cyber-attacks.
 
To address these issues, this paper presents a software architecture for enhancing safety, security, and predictability levels of learning-based autonomous systems. It leverages two isolated execution domains, one dedicated to the execution of neural networks under a rich operating system, which is deemed not trustworthy, and one responsible for running safety-critical functions, possibly under a different operating system capable of handling real-time constraints.
 
Both domains are hosted on the same computing platform and isolated through a type-1 real-time hypervisor enabling fast and predictable inter-domain communication to exchange real-time data.
The two domains cooperate to provide a fail-safe mechanism based on a safety monitor, which oversees the state of the system and switches to a simpler but safer backup module, hosted in the safety-critical domain, whenever its behavior is considered untrustworthy.
 
The effectiveness of the proposed architecture is illustrated by a set of experiments performed on two control systems: a Furuta pendulum and a rover. The results confirm the utility of the fall-back mechanism in preventing faults due to the learning component.

\end{abstract}

%\begin{keyword}
%Safe and secure machine learning \sep Reinforcement Learning \sep Simplex Architecture %\sep Functional Monitor

%\end{keyword}

%\end{frontmatter}

% \linenumbers

\section{Introduction}
\label{s:intro}

The recent explosion of artificial intelligence (AI) gave rise to the development of deep neural networks (DNNs) with super-human performance in many specific tasks, such as image classification~\cite{Masana22}, object detection~\cite{Zou23}, and control~\cite{Rybczak24}. 
These results attracted the interest of several industries, which started adopting deep learning methods in cyber-physical systems (CPSs), such as robots, drones, and vehicles, to perform complex tasks and increase their level of autonomy.
As a consequence, a growing number of autonomous systems are being deployed, and self-driving cars are seemingly a few years away.

Nonetheless, the outstanding performance of AI algorithms is not free from drawbacks.
One of the main obstacles to \textcolor{black}{the deployment of} deep learning models in CPSs is their low level of trustworthiness with respect to the strict industrial-grade standards applied in critical fields. Trustworthiness typically refers to a set of desirable properties that a learning-based CPS should have, including safety, security, and time predictability. 

\paragraph{Safety and fault-tolerance}
Deep learning models are known to produce imperfect predictions for many reasons. %\textcolor{black}{
A common cause is the distribution shift between train and test datasets that frequently occur when neural networks operate in the real world, e.g., when new unseen objects appear in the scene~\cite{quinonero2022dataset}.
%} 
In these cases, such unreliable outputs should be detected and excluded by a fault-tolerant mechanism, which should also prevent the propagation of faults and ensure that the system can be kept in a safe state.

\paragraph{Security}
Differently from the software development practice adopted for safety-critical systems, \textcolor{black}{the implementation of a neural network typically relies on sophisticated frameworks} (e.g,, PyTorch or TensorFlow) developed without following security- or safety-related programming practices, running on a rich operating system (e.g., Linux), and consisting of large code-bases that rely on several third-party libraries. These software components are all targets for cyber-attacks. For this reason, such components should be isolated from the safety-critical ones to protect the latter from possible attacks originated in non-trustworthy software. \textcolor{black}{In addition to that, DNNs are prone to several types of adversarial attacks~\cite{Sze14,Big18}, namely small malicious perturbations applied to the input that are imperceptible to humans, but can cause a neural network to produce a wrong output with a very high confidence score.}
Such adversarial attacks can also be performed in the physical world, i.e., without altering the digital representation of \textcolor{black}{the DNN input}, through dedicated patches or 3D-printed adversarial objects placed in the environment~\cite{Kur17,Nesti_2022_WACV}.

\paragraph{Time Predictability}
Popular software and hardware platforms used to execute DNN inference tasks typically exhibit low predictability in the time domain, which causes model outputs to be delivered with highly variable delays.
In fact, both deep learning frameworks and modern hardware acceleration technologies are designed to optimize the average-case behavior, rather than the worst-case one.
Consequently, response times are subject to considerable variability that can degrade the overall system performance and compromise system stability~\cite{Cas20-spe}.
For instance, accelerating neural models on general-purpose graphics processing units (GPGPUs) significantly reduces the average response time, but introduces highly variable delays in their execution~\cite{Cav17,But22}. This phenomenon is exacerbated if multiple DNNs are to be executed on the same accelerator, due to the higher number of conflicts that may arise on the shared computational resources (e.g., caches, buses, memory banks, etc.).

\subsection{This paper}
\textcolor{black}{To cope with these problems, this paper proposes an architectural solution that can address safety, security, and predictability issues in CPSs that include AI-based components, such as advanced sensory perception modules or neural controllers. }
This is obtained by \textbf{(1)} separating computations in two execution domains established by a real-time secure hypervisor and 
\textbf{(2)}: building a Simplex architecture driven by a \emph{safety monitor}. 

The two domains are defined as follows: a \textit{rich domain}, in charge of running all the learning-based components on a rich OS, and a \textit{safe domain}, responsible for executing all the safety-critical functions in a more protected environment. 
In our experimental setting, the two domains are hosted with strong isolation on a single AMD Zynq Ultrascale+ system-on-chip and managed by CLARE, a type-1 real-time hypervisor that is part of the Clare Software Stack~\cite{Clare}.
The two domains cooperate by means of inter-domain communication provided by the CLARE hypervisor, which guarantees bounded execution latency and jitter, as well as strong isolation mechanisms, such as secure cache partitioning and memory bandwidth reservation~\cite{Mod18} to reduce inter-domain interference. Security hardening of whole system is established at the hypervisor level by means of run-time security monitoring, address space layout randomization, and control-flow integrity.

To enhance safety, this work relies on the Simplex architecture~\cite{Sha94,Bak09,Moh13}, which relies on a \textit{safety monitor} that analyzes the state of the system, and/or the behavior of DNNs, commanding a switch to a simpler but safer controller \textcolor{black}{to recover from potentially unsafe states.}
% [CAMBIATO] whenever the output of high-performance controllers is either judged to be unreliable or is found to bring the system toward an unsafe state.
%In this way, the system can exploit the performance of AI components while enhancing robustness with classical control techniques or similar predictable algorithms.

% [RIMOSSO] In this work, the high-performance controllers are learning-enabled components based on neural networks running in the rich domain, whereas the safety monitor and the safe controllers are executed in the safe domain.

The effectiveness of the proposed architecture has been tested and evaluated using two control systems: a Furuta inverted pendulum and an autonomous vehicle based on the AgileX Scout Mini rover equipped with a camera and a LiDAR.
In both cases, the proposed architecture proved to be capable of managing the unsafe behavior of the 
AI-based components 
in a reasonable and effective way.

In summary, this work makes the following contributions:
\begin{itemize}
  \item It proposes a general architecture for AI-powered cyber-physical systems on a platform with two execution domains managed by a type-1 hypervisor that provides strong spatial and temporal isolation between them while guaranteeing safety, security, and predictability requirements by means of a Simplex-inspired switching mechanism.
  \item It presents two effective implementations of the safety monitor for managing risky situations and faults in the AI component adopted in the use cases.
\end{itemize}

\section{Background and related work}
\label{s:related}

%Related works:
%-	architectural solutions to enhance safety and security in autonomous systems
%   o	(confronta paper Cittadini)
%-	Simplex architecture / runtime monitoring
%-	Other applications
 
\textcolor{black}{The research concerning safe and secure architectures for AI-powered CPSs is wide and multi-faceted, as reported by several surveys \cite{pereira2020challenges, olowononi2020resilient}.} Given the scope of this paper, we focus on previous work using DNNs for autonomous systems.

A line of research within the safe and robust deep learning focuses on verifying and testing DNNs. 
\textcolor{black}{Verification methods aim at formally proving that certain properties hold for a DNN, while testing methods seek to find corner cases that cause a DNN to produce incorrect outputs.}
% [MODIFICATO] The former is a family of methods that aims at mathematically proving that a certain property holds for the DNN, while the objective of the latter is to find corner cases that yield incorrect outputs for the DNN. 
Despite the broad interest in such topics, they are not addressed in this paper since the focus is on architectural aspects of safety and security, while neural networks are considered untrustworthy by nature. The interested reader may refer to~\cite{huang2020survey} for an in-depth discussion.
The key concept of our work is that the AI components are considered untrustworthy by definition. Hence, it is necessary to introduce a dedicated architecture to isolate their unsafe behaviors.

\subsection{Hypervisor-based Architectures}

As the use of DNNs in CPSs is a relatively new topic, there is still no established approach for developing safe, secure, and predictable AI-powered CPSs. Currently, several architectural solutions have been proposed to handle the untrustworthy behavior of DNNs. 

A typical approach consists of executing the learning-based components and the safety-critical ones on different platforms \cite{reke2020self,liu2017autopilot,gutierrez2018towards}, communicating via the Robot Operating System (ROS) or custom channels. Such design requires multiple platforms with redundant resources and hardware components, making it unsuited for autonomous systems that sometimes present stringent size, weight, and power requirements.

Although other single-platform architectures have been proposed \cite{o2019f1,meier2015px4}, adopting hypervisor technology, to the best of our knowledge, is becoming the de-facto solution for supporting mixed-criticality applications.
\textcolor{black}{In fact, type-1 hypervisors are favored in mixed-criticality systems due to their small code base, which reduces both attack surface and execution overhead, and their direct hardware control, which enhances security, safety, and time predictability.} Typically, they also offer time-predictable latency, efficient inter-domain communication, and effective scheduling of virtual machines \cite{cittadini2023supporting}. 

Several research efforts have integrated Type-1 hypervisors in specific CPS applications. For instance, Klein et al.~\cite{klein2018formally} used seL4 to separate trusted and untrusted software in UAVs, ensuring security and isolation. Almeida and Prochazka~\cite{almeida2009safe} employed PikeOS for secure partitioning in spacecraft avionics. Craveiro et al.~\cite{craveiro2009flexible} utilized ARINC-653 standard in the AIR operating system for aerospace systems, focusing on IA-32 and Sparc architectures. Pérez and Gutiérrez~\cite{perez2017handling} implemented real-time communication in the Xstratum hypervisor, integrating ARINC-653 standard with support for DDS. Farrukh and West~\cite{farrukh2022flyos} developed a low-overhead hypervisor solution by providing timing guarantees using SCHED\_DEADLINE for Linux, though it has limitations for complex AI-based solutions.
Biondi et al. \cite{biondi2021sphere} proposed a hypervisor-based architecture for safety-critical systems, providing isolation, security, and real-time communication, while Cittadini et al.~\cite{cittadini2023supporting} presented a dual-domain architecture applied to UAVs.
These studies highlight the versatility and advantages of Type-1 hypervisors in various CPS applications. Building on such previous work, our paper focuses on implementing an architecture for safe use of DNNs in the wild. 
\textcolor{black}{The main strength and novelty of this paper is that the proposed architecture is general enough to serve as a blackprint for safe, secure, and predictable integration of learning-based components across diverse types of CPS. The hardware and software stack can be configured depending on the specific application requirements, as demonstrated by the two reported case studies (an inverted pendulum and an autonomous rover). Such two prototypes provide concrete guidelines that readers can adapt for their own applications.}

\subsection{The Simplex architecture}
The Simplex architecture was originally proposed to safely test newly-designed and complex controllers on safety-critical systems \cite{Sha94} and perform safe online control system updates \cite{Seto98}, \cite{Bak11}. Further developments employed the Simplex architecture to enhance the security \cite{mohan2013s3a} and safety \cite{Vivekanandan16} of CPSs.
It was also recently used to adopt deep learning in safety-critical applications \cite{Bio20-esl}, in a programming framework to ensure safety of ROS-based systems~\cite{desai2019soter}, and also to safely integrate reinforcement learning components in CPSs \cite{phan2020neural}. 

The main idea behind the Simplex architecture is the use of the following three interacting components:
\begin{itemize}
    \item A \textit{safe controller}: it is a simple controller designed to manage situations that might lead the system in an unstable or dangerous behavior. It typically performs simple halt procedures and/or safe operations defined by classical control techniques.
    \item A \textit{high-performance controller}: it is an untrustworthy controller, typically showing higher performance with respect to the safe controller but without any guarantee of safety. 
    \item A \textit{safety monitor} or \textit{runtime monitor}: it is a module responsible for deciding whether it is safe to let the high-performance controller regulate the system or it is better to switch to the safe controller.
\end{itemize}

The switching rule adopted by the safety monitor is tailored for the specific system and requires to be designed for each use case. 
\textcolor{black}{This work uses the Simplex architecture to increase safety when a learning-based component is adopted in the system, presenting the implementation details for two specific use cases that can be useful for the practitioner interested in developing such systems. }

\subsection{Safety monitor}
Many different runtime monitoring functions have been proposed in the literature. 
From a classical control perspective, a closed-loop system can be analyzed around a certain equilibrium with Lyapunov-based techniques to estimate the region of asymptotic attraction (ROA)~\cite{Zubov65}, a subset of the state space where the system is guaranteed to converge to the equilibrium. Estimating the ROA of the system regulated by the safe controller is vital to extract the boundaries that the state shall not overcome when controlled with the high-performance controller, since it will probably become unrecoverable by the safe controller. 
When both the plant and the safe controller are simple enough, Lyapunov-based techniques might be the safest options, as they always provide subsets of the actual ROA; it is also possible to optimize the maximum-volume ellipsoid that fits inside the ROA \cite{Seto99}. 
This approach has been used to safely control a Furuta inverted pendulum in Section~\ref{s:exp_furuta}. 
Alternatively, neural networks can be trained to provide an estimation of the ROA~\cite{Ferreira97,Ferreira99}, although it is advisable to somehow certify the behavior of the neural network before deployment in the safety monitor.

A different approach is based on what is called reachability analysis~\cite{Bak14}, which estimates the set of states that might be reached from the current state. This indicates whether the system is on the verge of instability, or, in general, close to an unsafe state set. However, such techniques can be resource-intensive and might not be feasible for the stringent memory and processing requirements of real-time systems. Moreover, they also present scalability limitations when dealing with complex high-dimensional systems. 

In some complex cases, it is difficult to define equations of a system that are inclusive of the unsafe situations that the system might encounter. \textcolor{black}{In these cases, it is possible to define a safety envelope, as we adopted to achieve safe control of the rover. The safety envelope can be configured to impose limitations to the maximum velocity, as well as to restrict the allowed position or distance from obstacles in close proximity.}
Additionally, it is possible to use monitoring techniques that analyze the behavior of the neural network:  
Uncertainty estimation~\cite{abdar2021review}, out-of-distribution detection~\cite{yang2024generalized}, or adversarial defenses~\cite{rossolini2023defending} might be used in combination to find the situations where the neural network does not provide safe outputs. These techniques are not considered for this paper, but future work will investigate them.

\section{Proposed architecture}
\label{s:method}
%\textcolor{red}{Tutta questa sezione va rivista}

%Method:
%-	Description of the architecture
%   o	Role of the hypervisor
%   o	Separation of the domains
%   o	Communication between the domains
%   o	Safety monitor and safety controller (not specific)

%This paper describes the implementation of an architecture that includes a subset of the features presented in the architecture described in Biondi et al. \cite{biondi2019safe}. 
%The system consists of two isolated domains managed by the CLARE-Hypervisor \cite{Clare}, a type-1 real-time hypervisor running on a Xilinx UltraScale+ platform, as depicted in Figure \ref{f:architecture}.
%As depicted in Figure \ref{f:architecture}, the system is composed of two isolated domains, which are hosted and isolated on the same platform, thanks to a type-1 real-time hypervisor called CLARE-Hypervisor. A more detailed explanation of the hardware and software specifications is provided in Section \ref{s:exp}, while this section describes the functional features, as depicted in Figure .

This section provides a high-level description of the architecture introduced in Section~\ref{s:intro}, focusing on its main components. The implementation details specific to the two considered use cases are described in Section~\ref{s:exp_furuta} and Section~\ref{s:exp_rover}, respectively.

\begin{figure}[htb!]
\centering
\includegraphics[width=12cm]{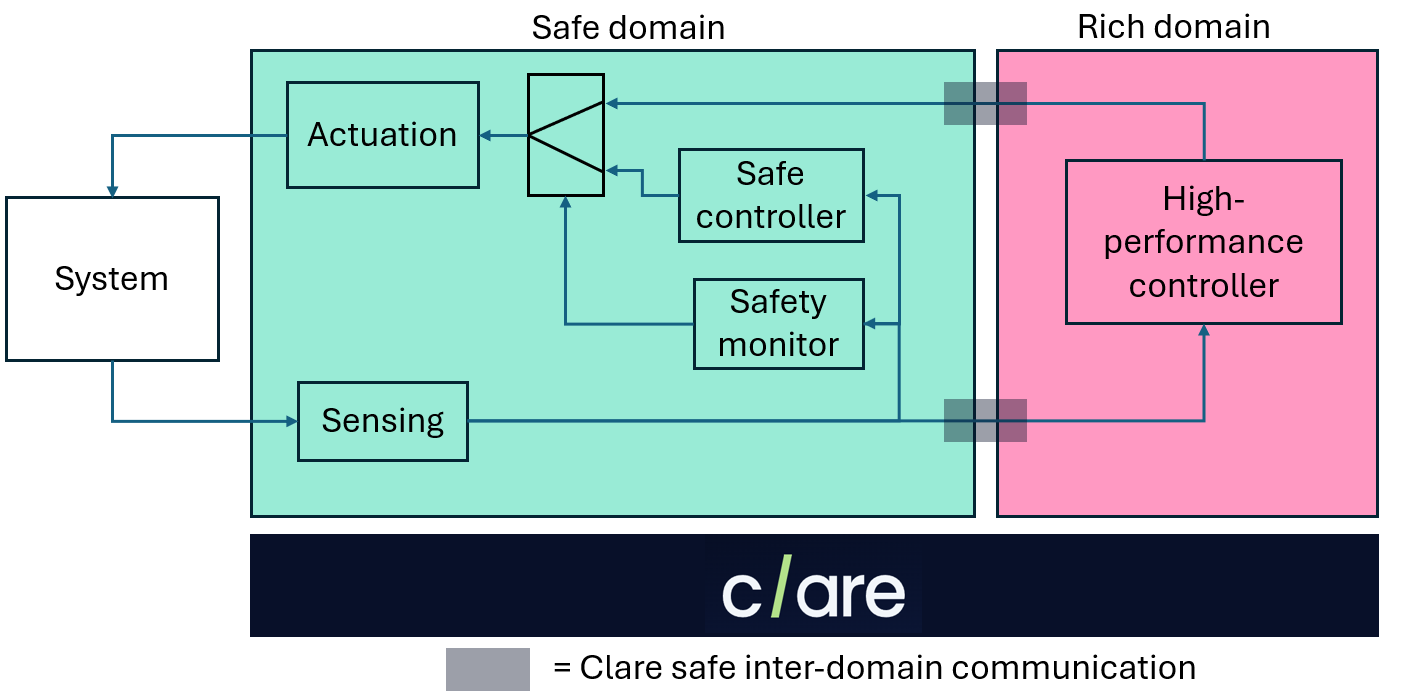}
\caption{Overall architecture of the system with functional blocks.}
\label{f:architecture}
\end{figure}

The proposed architecture is illustrated in Figure~\ref{f:architecture} and
consists of two execution domains concurrently running in strong isolation on the same hardware platform. CLARE-Hypervisor is in charge of \textcolor{black}{managing} two isolated Virtual Machines (VMs) on the hardware platform that may host stand-alone software execution environments powered by rich OSes (e.g., Linux) as well as real-time OSes or firmware (e.g., FreeRTOS). Each domain has both private and shared resources, whose access is defined at configuration time and then controlled by CLARE-Hypervisor at run time. In particular, each domain may have (i) exclusive access to computing resources (e.g., CPUs and accelerators), devices (e.g., Ethernet interfaces, DMAs), and memory resources, and (ii) mutual access to other physical and logical resources (e.g., shared-memory areas, inter-domain signals) that are mainly used for inter-domain communication mechanisms.
\textcolor{black}{The proposed architecture is composed of two isolated VMs that host the Safe and Rich domains, respectively.
% [ELIMINATO] to accomplish the same mission.
Each domain has private access to its dedicated resources required to implement its functionality, and to two inter-domain channels to implement full-duplex communication.}

The two domains have a different levels of criticality:
\begin{itemize}
    \item The safe domain has a high criticality and is responsible for all the safety-critical tasks, such as sensing, actuation, communication, safe control, and safety monitoring. This domain is typically powered by a real-time operating system (RTOS).
    % MM: Valutare se rifrasare visto il dual-Linux sul rover
    \item The rich domain has a low criticality and is responsible for all the high-performance computations. In the scheme illustrated in Figure~\ref{f:architecture}, the only learning-enabled component is the high-performance controller, but other controllers and low-criticality tasks might be present. This domain is deemed untrustworthy with respect to safety, security, and time predictability, and it is typically powered by a rich operating system such as Linux.
    % MM: Valutare se rifrasare visto il dual-Linux sul rover
\end{itemize}

%\textcolor{red}{Aggiungere altri dettagli per descrivere tutti i blocchi illustrati in Figura 1. Descrivere il meccanismo di switching ed evidenziarlo nella Figura 1.}
The default operational mode of the system uses the high-performance AI-based controller in the rich domain. 
In such an operational mode, the safety monitor constantly evaluates the state of the system to detect any anomaly or dangerous situation. Whenever this occurs, the AI-based controller is disconnected and the output of the safe controller is used instead. Please note that while this is not showed in Figure~\ref{f:architecture} for simplicity, the safety monitor may be triggered also by other conditions specific for each implementation. For instance, the safety monitor can read the output of the high-performance controller or its internal activations (in case a neural network is used).
The safety monitor switches back to the AI-based controller whenever it is deemed safe, functionally acting as a state machine, which is general enough to implement the required switching strategy for the system at hand. For instance, one could use the dead-zone approach, hysteresis switching, or blending control~\cite{zhu2015optimal}.

 \textcolor{black}{CLARE-Hypervisor enhances system security through VM runtime monitoring. When anomalous execution behaviors are detected within a virtual machine, the hypervisor can be configured to shut down the affected VM and restore a trusted backup. Since the monitoring mechanism operates outside the VM, it remains isolated and cannot be compromised by malicious software running inside the VM. Such security features are not the focus of this study as they serve as complementary security mechanisms.}

\textcolor{black}{This architecture is general enough to be seamlessly adapted and configured in different ways to match the system requirements.
The actual configuration of the architecture in terms of resource assignment depends on the target application and its criticality level.}
Specific configurations are discussed in detail in the following sections when addressing the two case studies.

The proposed architecture provides multiple benefits: (i) Possibility of executing a high-performance but untrustworthy controller on a physical system, safely handling sporadic wrong outputs by switching to a simpler but safer controller; 
(ii) Capability of safely managing deadline misses caused by the unpredictable timing behavior of the rich OS by exploiting the switching mechanism; (iii) Capability of isolating all the safety-critical components into the high-criticality domain, protecting the physical system from possible cyber-attacks to the rich OS or adversarial attacks against the AI components; \textcolor{black}{(iv) Run-time security monitoring at the hypervisor level;} (v) Possibility to train a neural controller directly on a physical system to fine-tune the network to refine the model.

To achieve the benefits mentioned above, however, the two isolated domains have to be carefully managed by properly addressing the following potential issues:
\begin{itemize}
    \item The use of two controllers in alternation could result in an unstable behavior if not properly handled. %\TODO{AB: this is a list of issues, while the next sentences are about solutions. They must be moved away.}
    \item The end-to-end latency resulting from the cooperation of the two domains must be precisely characterized and taken into account in the design of software activities since it could jeopardize the system stability or result in poor performance due to the reduced utilization of the AI components.
    \item \textcolor{black}{Typically, an embedded computing platform has several devices. Each device must be managed by a domain, potentially decreasing the performance of the system;} hence, the design of the architecture must be decided carefully, depending on the specific application requirements.
\end{itemize}

All these aspects are analyzed in the context of two use cases presented in Sections \ref{s:exp_furuta} and \ref{s:exp_rover}, each discussing a specific architectural configuration, the design of the control loop period, and the switching rules of the safety monitor.

\section{First case study: the Furuta pendulum}
\label{s:exp_furuta}

This section describes the specialization and implementation of the architecture presented in Section~\ref{s:method} to safely control a Quanser Furuta pendulum~\cite{Quanser}.
An overview of the hardware system and the software tasks is first provided and then followed by a description of the safety monitor and the controllers. Experiments are finally presented for evaluation purposes.

%Specifically, Section \ref{ss:exp_pendulum_system} introduces the system, Section \ref{ss:exp_pendulum_hwsw} presents the hardware configuration and the software tasks, Section \ref{ss:exp_furuta_controllers} details the finite state machine, the safety monitor and the controllers used in the experiments, and finally \ref{ss:exp_pendulum_results} reports the results of a sample run of the controlled system when perturbed with external disturbances. 
%

The primary safety requirement considered in this basic system is that the pendulum's pole must always be kept in equilibrium. As the dynamics of the pendulum evolves rapidly, this system constitutes a good case study to assess the real-time capabilities of the proposed control architecture. Please note that, although DNN-based controllers are not strictly required to control an inverted pendulum, this system provides interesting and valuable insights useful for the implementations on more complex systems, such as the one described in Section \ref{s:exp_rover}.

\begin{figure}[htb!]
\centering
\includegraphics[width=6cm, scale=\FIGSCALE]{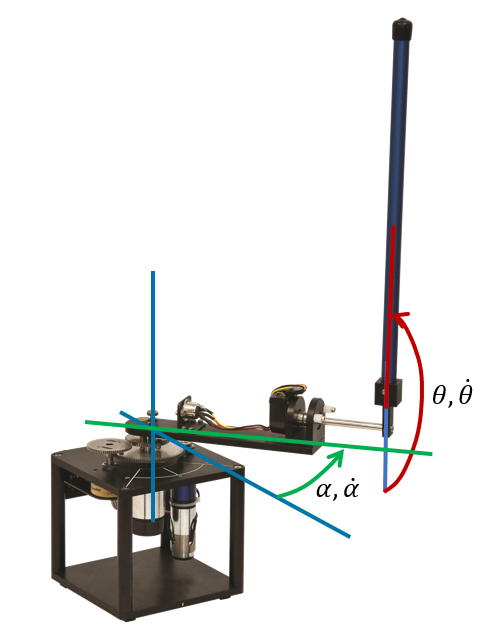}
\caption{Quanser's Furuta pendulum used in this demo.}
\label{f:pendulum}
\end{figure}

%\subsection{The system}\label{ss:exp_pendulum_system}
The Furuta pendulum, technically a rotary inverted pendulum, is in fact one of the standard benchmarks for real-time control problems.
Figure~\ref{f:pendulum} shows the pendulum used in this work.
The controller acts on the DC motor voltage with a PWM signal. The motor axis is linked to the horizontal arm (from now on simply indicated as the arm), while the pendulum rotates freely at the end of the arm. Both  angle values are acquired by an encoder. The angle of the arm is $\alpha$, while the angle of the pendulum with respect to the vertical is $\theta$.
The state of the system consists of the arm and pendulum angles and the corresponding angular velocities, formally $\x=(\theta, \alpha, \dot{\theta}, \dot{\alpha})^T\in \mathcal{X}\subset {\rm I\!R}^4$.  
The system is modeled in the standard form $\dot{\x}=f(\x, u)$, where $u\in {\rm I\!R}$ is the control input and $f$ is the non-linear dynamics.
The equations of the system 
can be found in Cazzolato et al.~\cite{cazzolato2011dynamics}.

The pole is brought in the upright position with a standard energy-based swing-up controller~\cite{astrom}, while the high-performance controller is used for pendulum stabilization when it is already up. 

% =================================================================
% HW/SW
% =================================================================
\subsection{Hardware and on-board software}
\label{ss:exp_pendulum_hwsw}

The reference computing platform is an AMD Zynq Ultrascale+ MPSoC (mounted on a ZCU102 board), a heterogeneous SoC that includes four Cortex-A53 cores, two Cortex-R5 cores, an FPGA-based Programmable Logic (PL), and several I/O peripherals~\cite{us_trm}. 
Thanks to its level of heterogeneity, this platform is suitable for different types of applications and it is also a good candidate for hosting multiple applications with mixed levels of criticality. 
The platform is managed by the CLARE Software Stack
%, which offers an abstraction layer of safe, secure, and interference-free virtual machines to host mixed-criticality applications on the same board across multiple domains. In particular, as previously mentioned, the system includes 
to establish two execution domains:
\begin{itemize}
    \item a \textit{Rich VM}, powered by Petalinux 2018.2 (an AMD/Xilinx Linux distribution), using three virtual CPUs with 2 GB of RAM and hosting the Caffe framework to perform the inference of neural networks;
    \item a \textit{Safe VM}, powered by the Erika3 RTOS~\cite{Erika}, using one virtual CPU with 256 MB of RAM, which hosts the safe controller, the system monitor, and a logging and telemetry component.
\end{itemize}

The sensing part is composed of two position encoders, one for the arm and one for the pendulum, which are sampled at 16 kHz through a specific logic synthesized on the FPGA. An Erika3 task controls the PWM actuation by driving a dedicated timer. 
Another Erika3 task in charge of telemetry sends the variables of interest through an UART port with a baudrate of 115200. The data are then sent to a web-based application running on a PC for monitoring and logging purposes.

\begin{figure}[htb!]
\centering
\includegraphics[width=155mm, scale=\FIGSCALE]{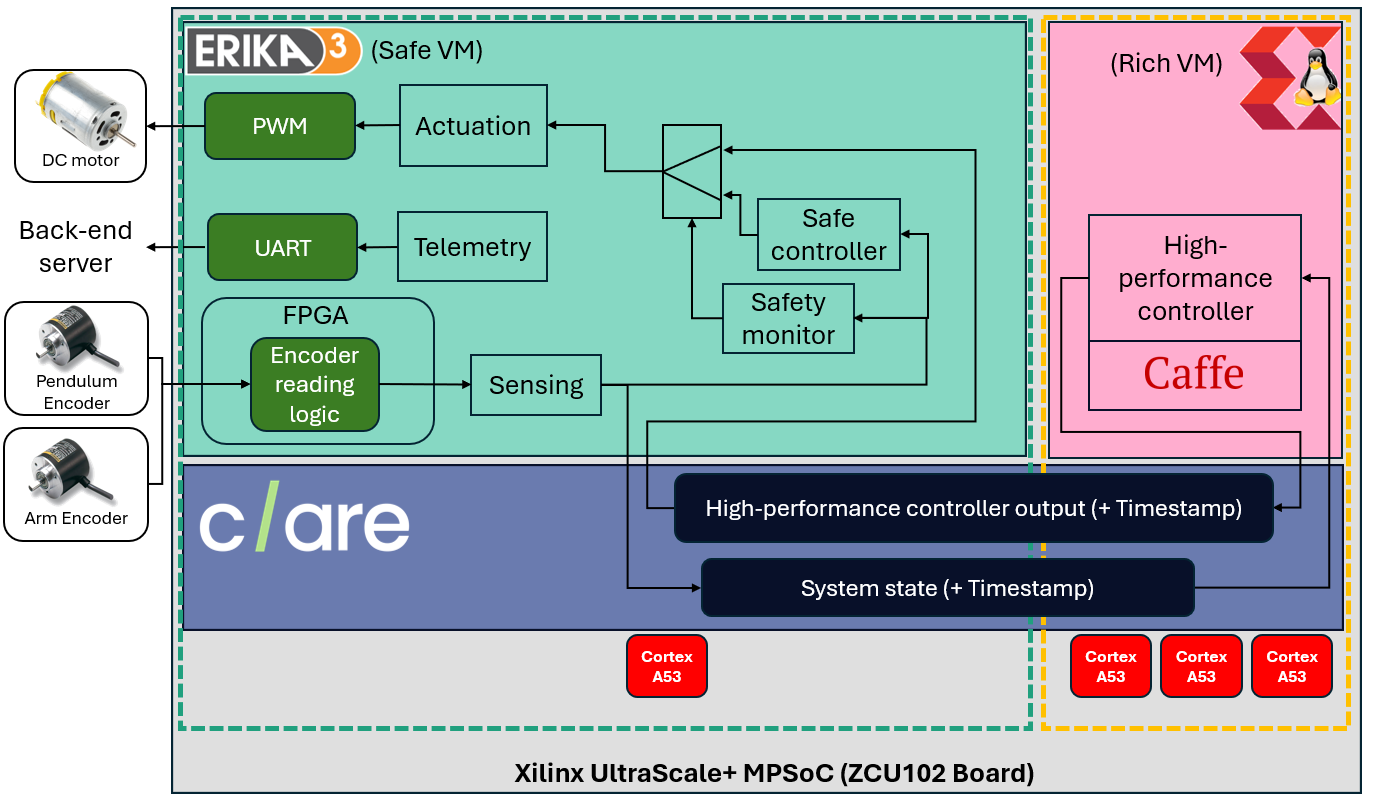}
\caption{Scheme of the hardware/software architecture managed by the CLARE-Hypervisor for safe and secure control of the Quanser rotary inverted pendulum. Please note that the blocks represent functional components and might not have a direct correspondence with the task set.}
\label{f:hw_sw_architecture}
\end{figure}

As shown in Figure~\ref{f:hw_sw_architecture}, the Rich VM has no direct access to any hardware device involved in the plant control loop.
Every access is mediated by the Safe VM, which communicates with the Rich VM through the inter-VM communication mechanism provided by CLARE. 
In particular, Erika3 tasks sample the position of both the arm and pendulum, computes the state of the system, and sends it to the Rich VM. The latter continuously monitors this port through a polling task, waiting for a new message. If the port contains a new message, a Linux application reads the state and performs a neural network inference using the Caffe framework. The output of this neural network is sent to the high-performance controller through another port and the loop is repeated.
In this way, the Safe VM can safely read the output of the high-performance controller from this dedicated port.
The sampling ports have a 1-to-N semantic and can contain one sample only. In this case, the channel is configured as a 1-to-1 port and always contains the freshest sample. 

The software running on Erika3 combines auto-generated code (from Matlab/Simulink) and hand-written code. The Erika3 task set is shown in Table~\ref{t:exp_rover_tasks}. In the priority column, higher values indicate higher priorities.

\begin{table}
\centering
 \begin{tabular}{||c c c c||}
 \hline
 \textbf{Task} & \textbf{Period [ms]} & \textbf{Deadline [ms]} & \textbf{Priority} \\ [0.5ex] 
 \hline
 \hline
 SafetyMonitor & 2 & 2 & 4 \\
 \hline
 Control & 4 & 4 & 3 \\
 \hline
 Stream & 10 & 10 & 2 \\
 \hline
 Interface & 20 & 20 & 1 \\
 \hline
\end{tabular}
\caption{Period, deadline and priority of the task set of the Safe VM (Erika3).}
\label{t:exp_rover_tasks}
\end{table}

The \textbf{SafetyMonitor} and the \textbf{Interface} tasks host the auto-generated code to handle the finite state machine required for the safety monitor switching mechanism. 
The safety monitor (i) checks the risk of instability according to the system state, and (ii) checks whether the output port of the high-performance controller contains fresh messages, meaning that Linux is not missing any deadline. 
The safety monitor algorithm and the other controller involved are described in Section~\ref{ss:exp_furuta_controllers}.

The \textbf{Control} task is responsible for:
\begin{itemize}
    \item sensing (acquiring measurements from encoders);
    \item sending the measurements to the state port (towards the rich VM);
    \item executing the safe controller algorithm;
    \item reading the advanced controller output from Linux through the dedicated port; and
    \item sending the correct actuation value to the PWM, according to the decision of the safety monitor.
\end{itemize}

Finally, the \textbf{Stream} task is responsible for periodically sending the system status to a web-based application through the UART, for visualization purposes.

%%%%%%%%%%%%%%%%%%%%%%%%%%%%%%%%%%%%%%%%%%%%%%%%%%%%%%%%%%%%%%%%%%%%%%%
% CONTROLLERS
%%%%%%%%%%%%%%%%%%%%%%%%%%%%%%%%%%%%%%%%%%%%%%%%%%%%%%%%%%%%%%%%%%%%%%%

\subsection{Controllers and safety monitor} \label{ss:exp_furuta_controllers}

Two controllers
are employed: the swing-up controller and the safe controller.
\textcolor{black}{A finite state machine 
decides which controller must be activated, depending on the state of the system.} The state machine runs on the safe domain, together with all the safety-critical tasks (i.e., the two controllers mentioned above and the safety monitor).

At the system start-up, the pendulum is assumed to stand in the down-hanging position $\x=(0, 0, 0, 0)$, from which a standard energy-based swing-up controller \cite{astrom} is used to bring the pole close to the upright position $\x=(0, \pi, 0, 0)$.
\textcolor{black}{The safe controller used for this application is a standard proportional controller $u=K\mathbf{e}$, where $\mathbf{e}$ is the error between a certain reference point and the current state $\x$, and $K \in {\rm I\!R}^{1\times 4}$ is computed by standard linearization and pole-placing techniques to guarantee a stable behavior.}

The high-performance controller is described in Section~\ref{ss:exp_pendulum_results}, but its behavior does not affect the switching mechanism of the safety monitor, which solely relies on the definition of safe controller and the notion of Region of Asymptotic Attraction (ROA).

Given an equilibrium point $\x_0$ in the state space (it can be chosen $\x_0=\mathbf{0}$ without loss of generality), the ROA $\mathcal{R}$ is the subset of the state space where the system is asymptotically attracted to the equilibrium, i.e., the system is stable. In this implementation on the Furuta pendulum, we are interested in estimating the ROA of the system controlled by the safe controller, since we want to know which states can be safely recovered to the equilibrium. %In this case, the concepts of safety and that of stability on the upright equilibrium overlap, since the objective is to keep the pendulum standing at all times. 

%\TODO{$\x(t)$ and $\mathcal{X}$ are undefined}  %FN: X è definito sopra
Formally, the ROA of the controlled system is defined as
\begin{equation}
    ROA = \{\x \in \mathcal{X} \quad s.t. \quad \lim\limits_{t\rightarrow\infty}\x(t) = \mathbf{0}, \x(0)=\x\},
\end{equation}
\noindent where $\x(t)$ is the state indexed in time $t$.

The ROA of a non-linear system cannot be computed in closed form, and must be estimated numerically. Section~\ref{s:related} provided a brief review of the methods proposed in the literature to estimate the ROA. The ROA estimation technique used in this paper is based on approximating the ROA as an ellipsoid centered in the equilibrium (i.e., the origin), having a quadratic form $\x P\x<1$, where $P\in {\rm I\!R}^{4\times 4}$. The value of the symmetric positive-definite matrix $P$ can be found with an optimization process that aims at maximizing the volume of the ellipsoid, as explained in~\cite{Seto99}.

By the ROA definition, it is possible to define a subset $\mathcal{R}$ of the state space that contains only the states that can be recovered safely to the equilibrium by the safe controller, that is
\begin{equation} \label{eq:ROA}
    \mathcal{R} = \{\x \;|\; \x^T P \x < 1\}\subset ROA.
\end{equation}

If the current state $\x$ can be checked to be in $\mathcal{R}$, then we can ensure that $\x$ is recoverable by the safe controller and the high-performance controller can be safely used (i.e., without the risk of instability). As soon as the current state exits $\mathcal{R}$, the safety monitor switches to the safe controller that is able to bring the state back to the equilibrium, thus preventing risky conditions that could possibly be caused by the high-performance controller.
This kind of switching between the high-performance and a safe controller is typical of the Simplex architecture~\cite{Sha94}, as it allows exploiting the higher performance of non-certified controllers safely.

To prevent frequent switches between the two controllers, the commutation to the high-performance controller occurs when a safe subset $\mathcal{S} \subset \mathcal{R}$
is reached by the state. In particular, the safe subset can be defined as
\begin{equation}
    \mathcal{S}(\Theta_s) = \{\x \;|\; \x^T P \x < \Theta_{s}\},
\end{equation}
where $\Theta_{s}<1$ is a user-selected threshold that defines how close to the equilibrium the safe subset should be.

\begin{figure}
\centering
\includegraphics[width=0.8\textwidth]{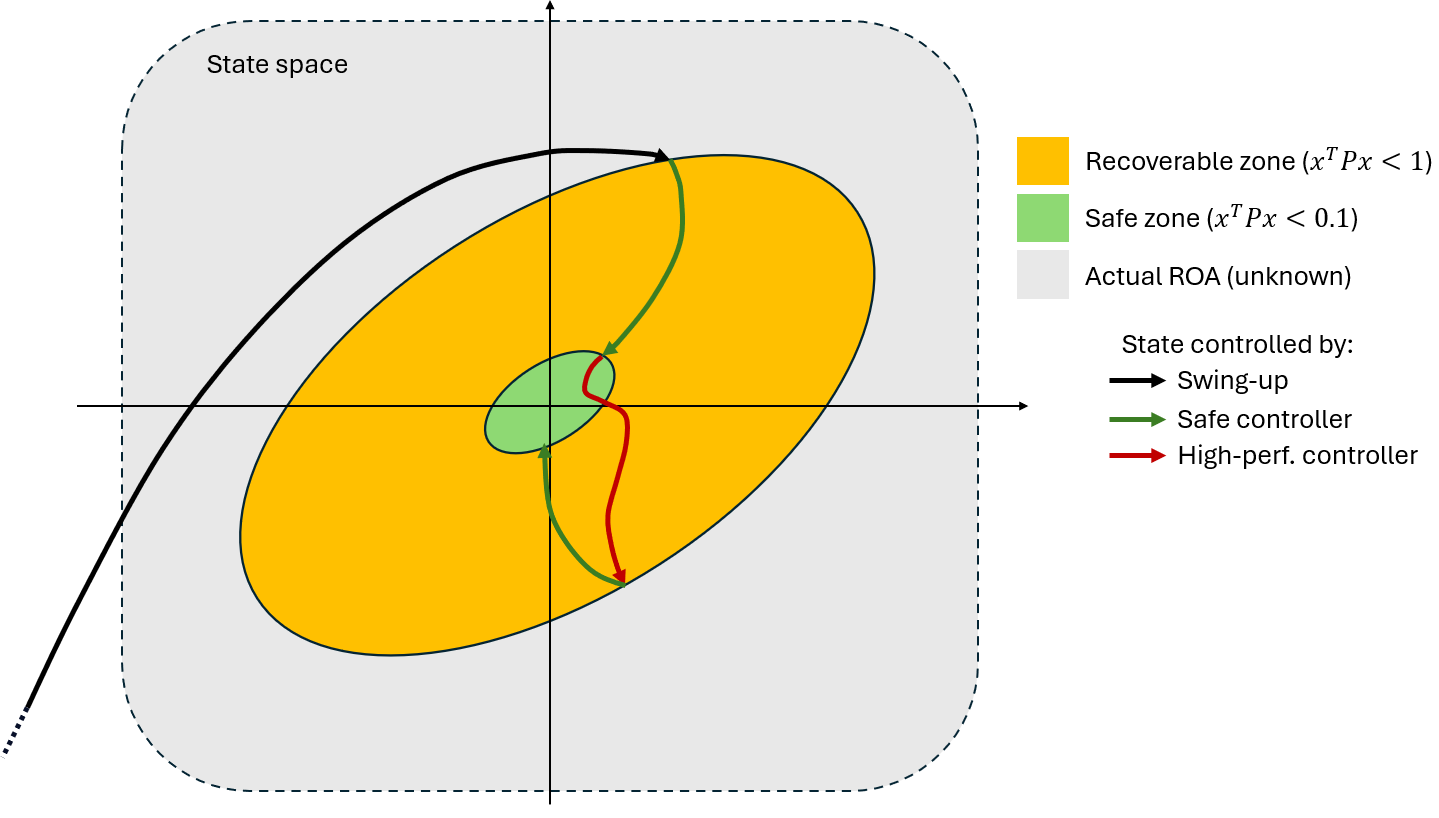}
\caption{ \label{f:lyap} Simplified 2D visualization of the state of the system and the various regions that define the behavior of the state machine. The actual region of asymptotic attraction (gray area) including all the states that are recoverable by the safe controller) is actually unknown. At system startup, the swing-up controller (black arrow) brings the state close to the equilibrium. As soon as the state enters $\mathcal{R}$ (yellow - states that can be surely recovered by the safe controller), the safe controller (green arrow) is activated to get the state close to the equilibrium and into $\mathcal{S}$ (green), where the high-performance controller can be activated safely. The state machine from this point switches between the high-performance and the safe controller depending on the instability index $x^TPx$.}
\end{figure}

Hence, the scalar quantity $\x^T P \x$ represents a sort of \textit{instability index}. When such an index is close to 0, it means that the state is close to the equilibrium. Conversely, when the instability index is close to 1, the system is close to an unrecoverable instability.
We empirically chose the threshold $\Theta_{s}=0.1$ to allow the safety controller to bring the state close to equilibrium before allowing the high-performance controller to intervene again.
A simplified visualization of the stability regions used by the switching-based state machine is illustrated in Figure~\ref{f:lyap}, which shows the actual ROA of recoverable state (gray area), which is usually unknown, the estimated ROA $\mathcal{R}$ (yellow area) and the safe region $\mathcal{S}$ (green area).

%Please note that the state space lies in ${\rm I\!R}^4$, while this figure shows a simplified 2D state for visualization purposes. 

It is worth noting that to guarantee the stability of the overall system, a one-step-ahead simulation of the state is required. This is because, in general, a wrong control input at time $t$ issued by the high-performance controller could move the state $\x(t+1)$ not only outside the ROA approximation $\mathcal{R}$, but also outside the actual, unknown ROA. To prevent this to happen, the safety monitor must work on a prediction $\hat{\x}(t+1)$ of the system state rather than on the actual state $\x(t)$. However, because of (i) the strict timing and resource constraints of the system, and (ii) the fact that $\mathcal{R}$ is typically a coarse under-approximation of the ROA, this possibility was ignored, also because never occurred during our extensive experiments. Nevertheless, it should be taken into account in the design of such an architecture.

\textcolor{black}{It is also worth remarking that all the ROA estimation methods might suffer from false negatives. Compared with false positives, false negatives are the most dangerous threat for the safety monitor. In fact, classifying an unsafe state as safe might lead to catastrophic failures. 
There is no closed-form algorithm to understand whether false positives exist. Simulation can however be used to sample the boundary of the estimated ROA to count the number of false positives (which might become unfeasible for highly-dimensional systems). However, the simulated model includes both parametric uncertainty and non-linear modeling inaccuracies with respect to the real-world version. Hence, in our specific case, the best option is to sample the estimated ROA boundary multiple times, each time varying the numerical parameters of the model. With a 5\% variation and a grid size of 0.05 in the state space we found no false positives, which confirms that the ellipsoid approximation is a coarse under-approximation of the actual ROA (although such a sampling cannot provide an absolute guarantee). If false positives were to be found, it might be useful to model a variable $\Theta_U\in(0, 1)$ and define a parametric version of the estimated ROA: $\mathcal{R}(\Theta_U) = \{\x \;|\; \x^T P \x < \Theta_U\}\subset ROA$.}

%\subsubsection{Advanced Controller}

\subsection{Results} \label{ss:exp_pendulum_results}

To carry out experiments on this case study, the high-performance controller was implemented by a shallow neural network (one hidden layer, with tanh activation) taking the state $\x$ as input and returning the scalar output $u$ required to control the system. The hidden layer has $H=128$ neurons, and the network was trained by the Covariance Matrix Adaptation (CMA-ES)~\cite{auger2012tutorial}) evolutionary algorithm. 
The optimization process was performed in episodes, as in a standard reinforcement learning framework. An episode ends either after 1000 steps (with sample time 4 ms) or by being interrupted as soon as the pole angle $\alpha$ gets 20 degrees off the vertical or the arm angle $\theta$ drifts 45 degrees off the starting point.
At each episode instant, the reward is
$1 - 0.1 |u|$,
%-------------------------------------
%GB: Il reward non dipende dai fallimenti (alpha > 20 o tetha > 45)? Spiegare meglio.
%-------------------------------------
% FN: non serve punire per i fallimenti, se l'episodio finisce presto la reward è già bassa. Ho aggiunto un commento.
which is summed until termination of the episode, similarly to what happens in the CartPole OpenAI-gym~\cite{brockman2016gym} environment. The term proportional to $|u|$ serves to enforce regularization. During training, random perturbations were injected to make the agents learn to withstand small pushes.
The optimization was performed in a simulated environment, specifically a custom OpenAI-Gym environment that replicates the discrete dynamics of the Furuta pendulum.

\begin{figure}
\centering
\includegraphics[scale=0.4]{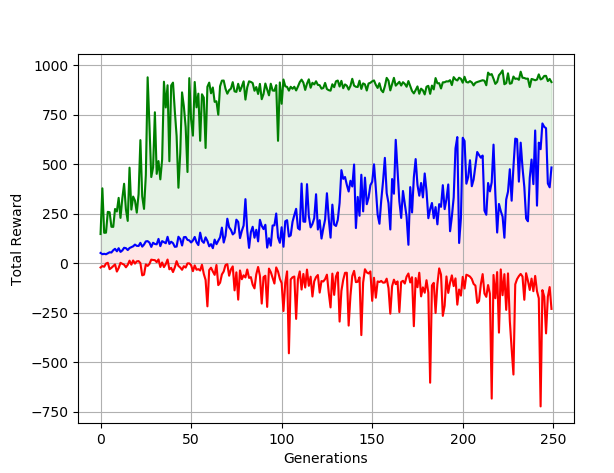}
\caption{Results of the training process for the population using CMA-ES. Green is the best for each generation, black is the median, red is the worst; shaded areas include all the individuals in between.}
\label{f:training}
\end{figure}

CMA-ES considers the set of parameters of the network as the chromosomes to be evolved. Since the neural network has 4 inputs, the hidden layer has $4\times H = 512$ weights and 128 biases, whereas the second layer consists of a single neuron with $H = 128$ weights and one bias. Hence, the total number of parameters is 769. The optimization was performed for 250 generations, with a population of 50 individuals, and a survival rate of 50\%. To have a smoother control on the real system, the action is averaged in a moving window of two samples.
Figure~\ref{f:training} shows the results of the training process, where it is clear that the best agents are consistently performing well after  just 100 generations.

\begin{figure}
\centering
\includegraphics[width=\columnwidth]{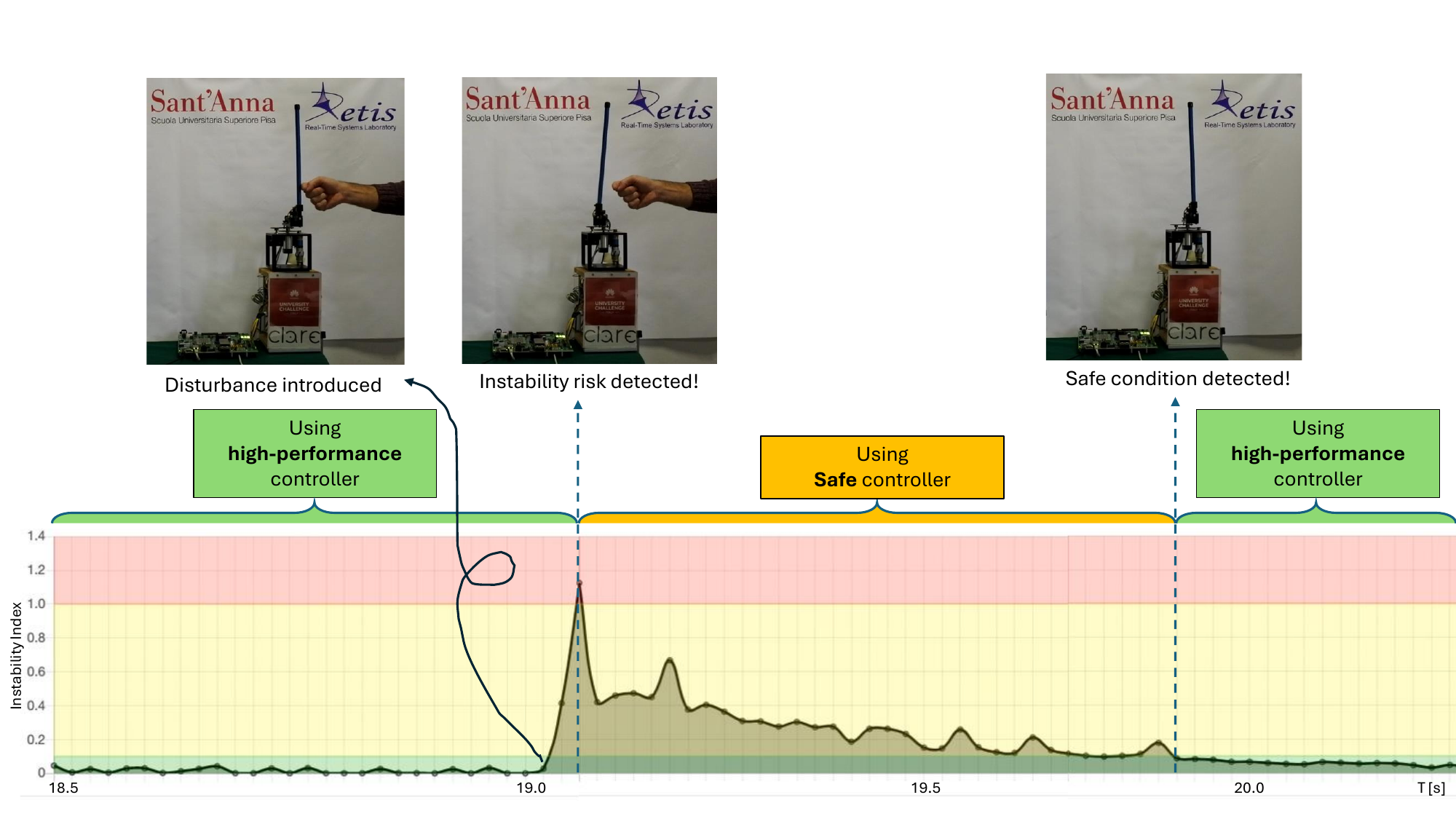}
\caption{A sample of the switching behavior of the overall architecture. When a strong disturbance is introduced, the high-performance controller brings the state of the system towards instability. This condition is detected by the safety monitor that promptly switches to the safe controller, which brings the state back to a safe neighborhood. The high-performance controller is then reactivated.}
\label{f:pendulum_run}
\end{figure}

Figure~\ref{f:pendulum_run} illustrates a situation in which the safety monitor decided to switch to the safe controller as a consequence of the injection of a disturbance (i.e., a push). A video\footnote{\url{https://www.youtube.com/watch?v=pcrC1NZ12L0&t=191s}} is also available on line to view the actual behavior.

Table \ref{t:exp_pend_times} shows the minimum and maximum time required for the execution of the different portions of the Control and Safety monitor tasks in 10.000 samples. The end-to-end latency of the control task is also showed (Linux2Erika end-to-end). This row in particular takes into account also the neural network inference time and the communication overhead between the safe VM and rich VM and back. It is possible to note that the maximum time required is less than the control task period (4ms).
%--------------------------------------
%GB: Per ora la lascerei. Poi vediamo. %--------------------------------------
\begin{table}[]
\centering
\begin{tabular}{l|l|l|}
\cline{2-3}
                                                   & Min {[}$\mu$s{]} & Max {[}$\mu$s{]} \\ \hline
\multicolumn{1}{|l|}{Sensing}                      & 0.789            & 2.387            \\ \hline
\multicolumn{1}{|l|}{Reading hp controller output} & 1.694            & 3.179            \\ \hline
\multicolumn{1}{|l|}{Compute instability index}    & 0.174            & 0.382            \\ \hline
\multicolumn{1}{|l|}{Finite state machine}         & 0.113            & 3.124            \\ \hline
\multicolumn{1}{|l|}{Writing hp controller input}  & 1.697            & 3.693            \\ \hline
\multicolumn{1}{|l|}{Linux2Erika end-to-end}       & 288.579          & 3975.657         \\ \hline
\end{tabular}
\caption{Min and max times}
\label{t:exp_pend_times}
\end{table}
\section{Second case study: the AgileX Scout Mini rover}
\label{s:exp_rover}

This section describes the implementation details of the architecture presented in Section~\ref{s:method} to control an AgileX Scout Mini rover~\cite{AgileX} for the task of LiDAR- and camera-based navigation and obstacle avoidance. 
Section~\ref{ss:rover_system} illustrates the rover and its features, Section \ref{ss:rover_hwsw} presents the specific architecture designed for this system, Section~\ref{ss:rover_controllers} specifies the controllers and the safety monitor used for the selected task, and Section~\ref{ss:rover_results} reports the results of the experiments performed on the physical system.

\begin{figure}[h!]
    \centering
    \includegraphics[width=0.8\textwidth, scale=\FIGSCALE]{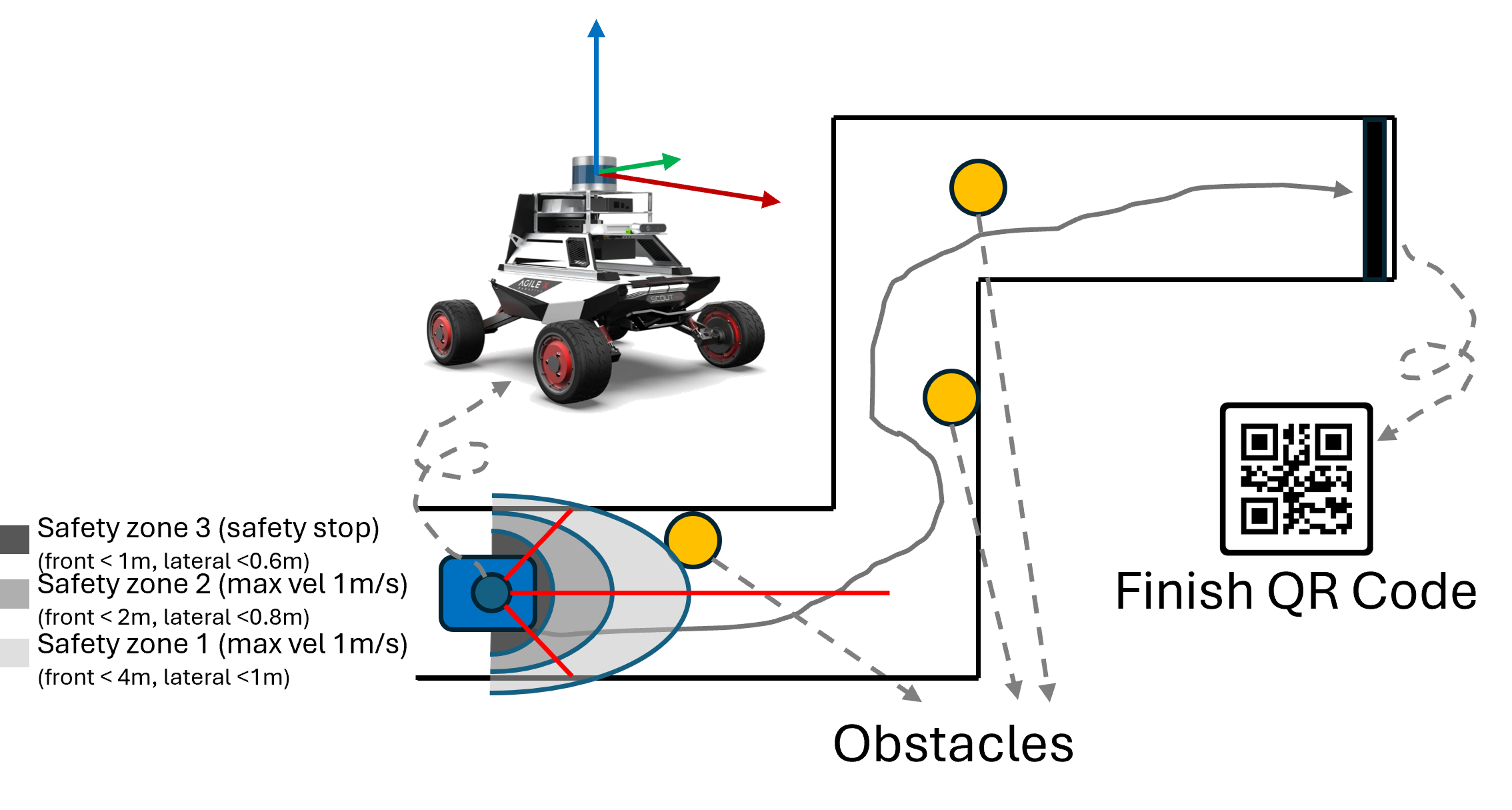}
\caption{\label{f:rover_system} \textcolor{black}{The AgileX Scout Mini rover used for the second use case and a scheme of the corridor navigation task performed. The picture also illustrates the LiDAR reference frame on the rover, the three LiDAR distances that constitute the output of the LiDAR pre-processing algorithm, and the three safety zones defined by the safety monitor.}}
\end{figure}

\subsection{The system}\label{ss:rover_system}

The AgileX Scout Mini rover is an agile and powerful platform, ideal for testing software for autonomous systems. Powered by a large battery, the rover is equipped with both a 
%Velodyne VLP-16 LiDAR and a RealSense D435 
LiDAR and a camera. Its design allows users to control it with built-in nodes that implement differential drive, making it behave like a unicycle.

The unicycle is a standard dynamical system whose state is described by $\x=(x, y, \theta)$, where $(x, y)$ represents its position on the plane and $\theta$ its orientation. 
Being a non-linear system, its dynamics can be described as $\dot{\x}=f(\x, \uu)$. However, differently from the pendulum, where $u$ was a single scalar, here $\uu=(v, \omega)$ is a two-dimensional vector where $v$ and $\omega$ are the linear and angular velocities (i.e., the yaw rate) of the rover, respectively.

The task carried out for this test case is navigation and obstacle avoidance, performed on a corridor using a LiDAR and a camera, as depicted in Figure~\ref{f:rover_system}. Specifically, the rover starts at a random point of the corridor and must arrive at the end point without colliding with the corridor walls. Additionally, a few obstacles are placed throughout the route and a QR code is placed at the end of the corridor to signal the end of the course. \textcolor{black}{The safety zones referred in the caption are introduced in Section \ref{ss:rover_controllers}.}

\subsection{Hardware and on-board software}\label{ss:rover_hwsw}

%The rover is shipped with a \textcolor{red}{Velodyne VLP-16} LiDAR and a RealSense camera. %, and an NVidia Jetson board \TODO{add specific jetson board}.

We setup the rover with a Velodyne VLP-16 LiDAR, a Logitech 920 camera, and an AMD Kria KR260 board, equipped with an AMD Zynq UltraScale+ MPSoC. \textcolor{black}{We chose the Kria KR260 board for two reasons: 
%MM: Si potrebbe dire "to reduce the SWaP impact (size, weight, and power)"
%    A quel punto "Compacr footprint" e "Low power consumption" si tolgono e sio lascia solo la parte dopo il :
\begin{itemize}
    \item Compact footprint: mounting the board on the Agilex Scout MiniRover required a board with a limited size.
    \item Low power consumption: definitely lower compared to other alternatives such as Nvidia boards \cite{lizano2024comparison}.
\end{itemize}
The selected board includes an FPGA device, which was not used in this particular setup. Nonetheless, it has the potential to accelerate and offload the inference of DNNs, which will be investigated in a future work.}

%Similarly to the ZCU102 introduced in Section~\ref{s:exp_furuta}, the board is equipped with different I/O peripherals including: 
%\begin{itemize}
%    \item Two USB 3.0 hubs: \texttt{USB0}, \texttt{USB1}.
%    \item Two Ethernet interfaces: \texttt{Eth0}, \texttt{Eth1}.
    % \item A Micro-USB UART/JTAG cable used for debugging.
    % \item A display port (\texttt{DP}).
%\end{itemize}

% details of SoC and board
% description of the VMs (particular emphasis on why the safe linux can be considered safe)
% use of ROS2 (and custom transport/Clare interface)
% Description of the architecture
% Specification of the tasks/nodes with deadlines.

% Explain the two domains and their tasks

\begin{figure}
    \centering
    \includegraphics[width=0.85\textwidth]{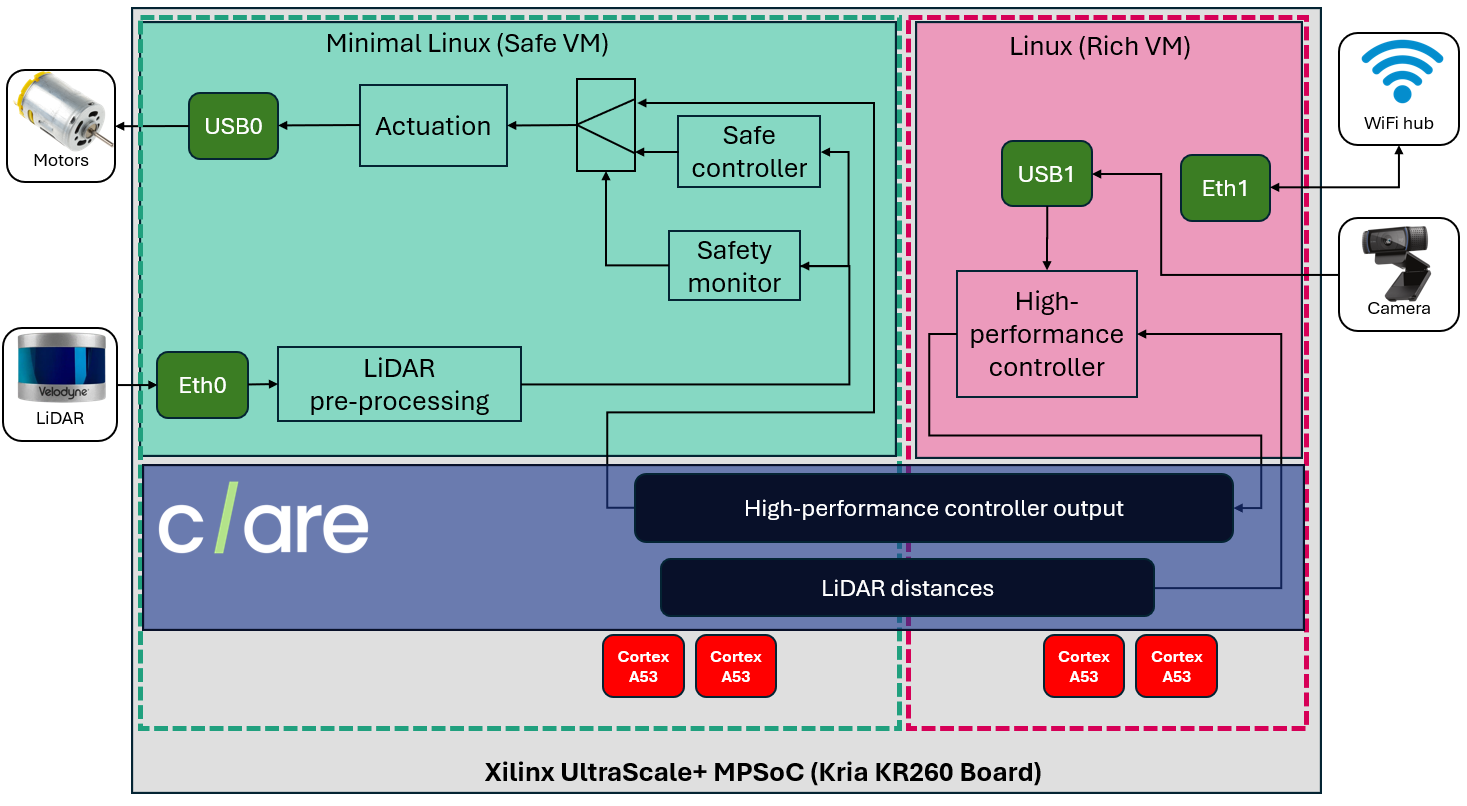}
    \caption{The hardware/software architecture managed by the CLARE-Hypervisor for safe and secure control of the AgileX rover.}
    \label{f:rover_architecture}
\end{figure}

Figure~\ref{f:rover_architecture} shows the hardware/software architecture of the system. As in the previous setup, platform virtualization is managed by CLARE. 
Since the application tasks present different criticality levels, two separate VMs running Ubuntu 22.04 LTS are utilized. 
They will be referred to as \texttt{Safe\_VM} and \texttt{Rich\_VM}. \textcolor{black}{The hypervisor has been configured to provide static partitioning of the hardware resources among the two VMs. In particular, CPUs and memory-mapped areas (DRAM and device registers) are exclusively assigned to the VMs. This strategy minimizes resource contention and hypervisor intervention in virtualization of the resource, thus also reducing the power consumption.}

Linux was chosen as the operating system for this dual-VM configuration due to its flexibility in supporting the execution of modern applications and for its wide support for third-party software, with particular reference to device drivers for cameras and LiDARs. Furthermore, Linux enables the execution of the ROS2~\cite{ros2} middleware, which is particularly useful to implement the tasks to be performed by the rover. ROS2 was used both as a communication infrastructure and to leverage pre-compiled sensing and actuation nodes controlling the velocity of the rover and collecting the point cloud data from the LiDAR. The Linux version running in the \texttt{Safe\_VM} was restricted to a minimal configuration, stripped of Internet access and all the redundant services that might interfere with safety-critical tasks. Conversely, the Linux version running in the \texttt{Rich\_VM} was kept with a stock configuration, including Internet access to enable remote access to non-critical services, which was also useful for simulating a cyber-attack (see Section~\ref{ss:rover_results}).

\textcolor{black}{Both the \texttt{Safe\_VM} or the \texttt{Rich\_VM} were exclusively assigned one of the two Ethernet ports (\texttt{Eth0} and \texttt{Eth1}). The same applies for the two USB 3.0 ports (\texttt{USB0} and \texttt{USB1}).}
The responsibilities of the two domains, corresponding to the two VMs, are summarized below:

\begin{itemize}
    \item Safe Domain (\texttt{Safe\_VM}):
    \begin{itemize}
        \item Capture the 3D point cloud from the LiDAR device through the \texttt{Eth0} interface and preprocess it to obtain distance measures from the LiDAR.
        \item Send the LiDAR distance measures to the \texttt{Rich\_VM} and receive the corresponding neural network output.
        %\item Implement the secure controller for the actuation based on the LiDAR distances. 
        \item Run the safety monitor, which enables actuation via either the safe or the high-performance controller output from the \texttt{Rich\_VM}.
        \item Manage actuation through an USB2CAN interface using the \texttt{USB0} interface connected to the CAN interface of the motors. 
                
    \end{itemize}
    \item Rich Domain (\texttt{Rich\_VM}):
        \begin{itemize}
        \item Read the LiDAR distance measures from the \texttt{Safe\_VM}. 
        \item Capture the image data from the Camera connected to the \texttt{USB1} interface.
        \item Execute the high-performance controller based on LiDAR distance measures and the RGB camera data and send the control to the \texttt{Safe\_VM}. 
        \item Implement a wireless communication using the \texttt{Eth1} interface.
    \end{itemize}
\end{itemize}

% % Safety is key (Already discussed in the next section)
% The safety monitor is the key component for preventing collisions with obstacles in the path. The high-performance controller is based on neural 
% Talk about Clare middleware 
The CLARE software stack provides a component, named CLARE-Middleware, that enables inter-domain communication using Cyclic Asynchronous Buffers (CABs) implemented in shared memory. A CAB is a data structure that operates using two pointers. The write pointer indicates where the data element will be written, while the read pointer shows where the next element will be read. Domains can write or read from these buffers according to the access control defined on the hypervisor configuration. To separate safe and unsafe data, the hypervisor is configured with two distinct buffers, referred to as \texttt{Safe2Rich} and \texttt{Rich2Safe}, to enable unidirectional communication between the domains. The \texttt{Safe2Rich} buffer is write-only for the \texttt{Safe\_VM} and read-only for the \texttt{Rich\_VM}, while the \texttt{Rich2Safe} buffer has the opposite access permissions.
% ROS2

\begin{figure}
    \centering
    \includegraphics[width=0.7\textwidth]{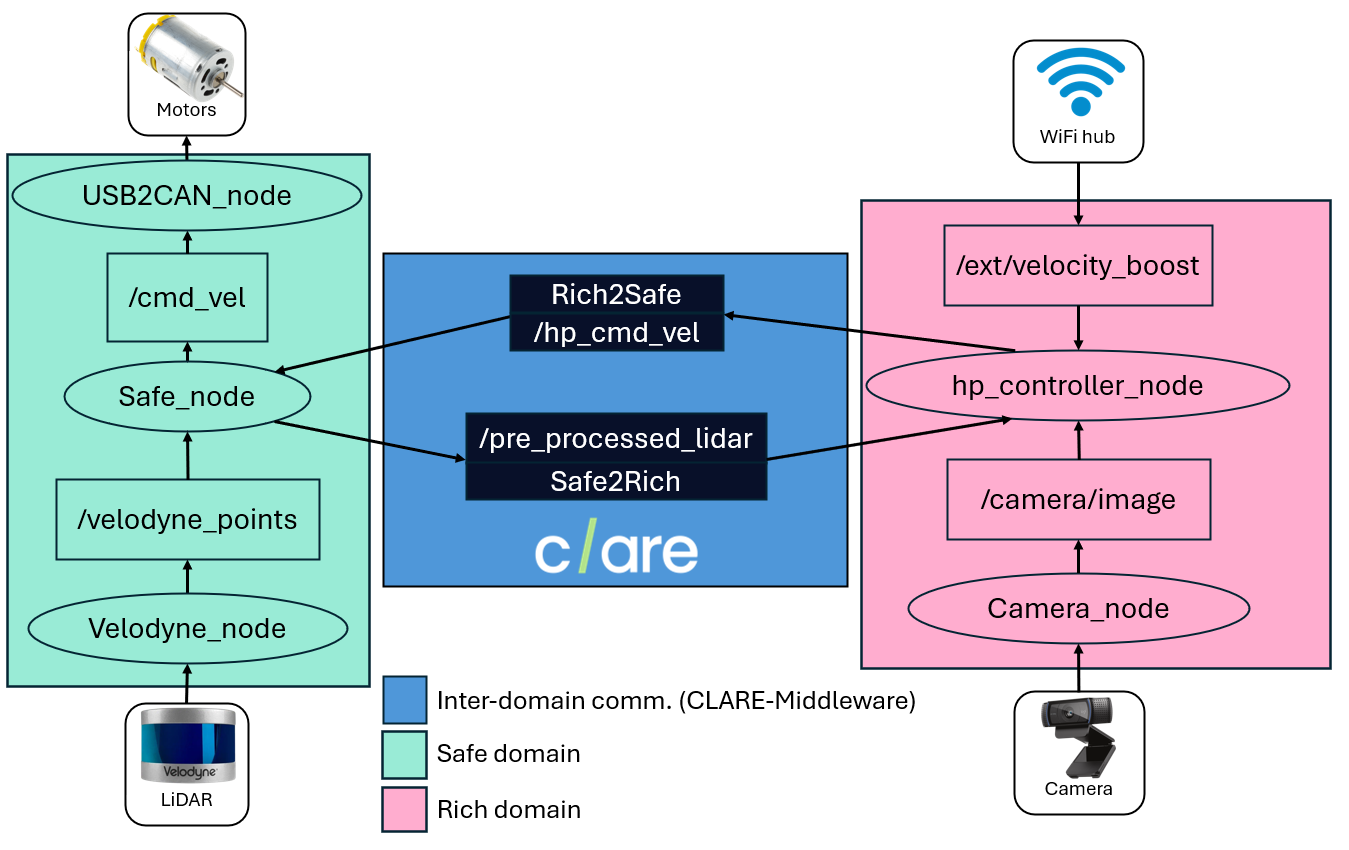}
    \caption{Graph of ROS2 nodes and topics. The inter-domain communication is implemented with a custom transport.}
    \label{f:rover_ros2}
\end{figure}

The two domains run distinct instances of the ROS2 Humble framework. The tasks are therefore organized in nodes and exchange information via ROS2 topics. Figure \ref{f:rover_ros2} illustrates how the nodes and topics interact. The \texttt{velodyne\_node}, the \texttt{USB2CAN\_node}, and the \texttt{camera\_node} are pre-compiled and provided by the vendors. The \texttt{safe\_node} implements all the functionality of the \texttt{Safe\_VM}: LiDAR pre-processing, safety monitor, and actuation. Conversely, the \texttt{hp\_controller\_node} runs the high-performance controller, consisting of a neural network implemented in C++ and performs the inference on the pre-processed the LiDAR data. The node also implements a QR code reader. The communication between the \texttt{safe\_node} and the \texttt{hp\_controller\_node} is managed by  CLARE-Middleware.
The \texttt{safe\_node} invokes the CLARE-Middleware API to write the pre-processed LiDAR data to the \texttt{Safe2Rich} buffer. The \texttt{hp\_controller\_node} polls the buffer for the data, processes it using the neural network, and sends the output to the \texttt{safe\_node} via the \texttt{Rich2Safe} buffer. The \texttt{safe\_node} polls the \texttt{Rich2Safe} buffer, executes the safety monitor, and applies the motor actuation (details in Section~\ref{ss:rover_controllers}). The polling is supervised by a timer to guarantee that the task deadline is respected.

% OLD
% Our ROS2 implementation is equipped with a custom FastDDS transport leveraging the CLARE middleware to enable communication between the two VMs. When a ROS2 node subscribes to or publishes a topic located in the other VM, the custom FastDDS transport acts as an intermediary, implementing data exchange via the \texttt{Safe2Rich} and \texttt{Rich2Safe} buffers. To ensure system safety, the visibility of topics related to critical tasks is restricted to nodes within the safe domain. For instance, the \texttt{cmd\_vel} topic, used for motor actuation control, is accessible only to the safety monitor. This restriction is essential to prevent the rich domain from bypassing the safety monitor and directly controlling motor actuation. The next section describes the technical details and rationale behind the design of the controllers and the safety monitor.

From a temporal standpoint, the LiDAR acquisition is the bottleneck, with a typical rate of 10Hz. Hence, all the nodes are configured to run with a period of 100ms.

Note that both the safe controller and the safety monitor are executed in the safe domain only and do not rely on camera inputs, but only on LiDAR data. 
In fact, images often suffer from several limitations, such as lighting conditions, perspective distortion, or other environmental factors. Furthermore, distance estimation from cameras, which is essential for obstacle detection and avoidance, is subject to higher errors with respect to LiDARs, even when considering stereo vision systems.
Conversely, LiDAR data is much more structured and directly provide precise distance measurements. Furthermore, certified LiDAR technologies are also available and do not necessarily require learning-enabled components, for which it is hard to achieve a high level of assurance, to process the data they produce. For these reasons, camera data are processed by the high-performance controller only, while LiDAR data are processed by the safe one to enable trustworthy perception for safety-related purposes.

The raw LiDAR data consists of a 3D point cloud, whose size depends on the type of LiDAR and the number of channels (i.e., vertical beams) it generates. For instance, the Velodyne VLP-16 returns up to 300.000 points every 100 ms.
However, since the specific obstacle avoidance application does not require such a massive amount of spatial data, the point cloud is first pre-processed to discard the points that are above the height of the rover (0.5 m). The selected points are then converted to distances from the LiDAR and grouped in bins depending on the angular position. The output of the LiDAR pre-processing is then a 3-dimensional vector $d$, where each element corresponds to the distance at -45, 0, and 45 degrees with respect to the x-axis of the LiDAR. This is because, in our case study, the rover was allowed to move forward only. %, as depicted in Figure~\ref{f:safety_zones}.

\subsection{Controllers and safety monitor}\label{ss:rover_controllers}

The safe controller was designed to limit the velocity of the rover to avoid collisions. The velocity limits are imposed according to the risk evaluated by the safety monitor.
The safety monitor defines three different safety zones with increasingly stringent speed limits. Outside of these zones there is no risk of collision, and the output of the high-performance controller is used directly.
The first (Z1), second (Z2), and third (Z3) safety zones are defined according to the distances returned by the front and lateral rays, as depicted in Figure~\ref{f:rover_system}. In Z1 and Z2, the Safe controller saturates the velocity command to 1.5 and 1 m/s, respectively. These numerical values depend on the rover breaking dynamics and were devised experimentally because of the high slip and slide motion given by the low friction between the rover wheels and the corridor pavement. The effectiveness of the zones is validated in the experimental subsection to avoid collisions when the rover is commanded to travel at its maximum velocity, which is about 3 m/s. Inside zone Z3, a safety stop is triggered by the Safe controller and the human operator must acknowledge it before re-enabling the high-performance controller. 
%high speed.

%\begin{figure}
%    \centering
%    \includegraphics[width=0.6\linewidth]{images/rover_imgs/safetyzones.png}
%    \caption{Illustration of the Safety Zones that define the behavior of the Safe Controller, which limits the maximum velocity allowed. }
%    \label{f:safety_zones}
%\end{figure}

The high-performance controller is a 4-layer neural network. The input layer consists of an array of three elements encoding the distances coming from the LiDAR preprocessing $d$. The two hidden layers are fully connected, have 100 neurons each, and use a ReLU-activation function, while the output layer has two fully connected neurons using the \emph{tanh} activation; the two output neurons form a vector $A(d)$ whose elements represent the rover control variables $(v, \omega)$. 
During regular operation, the output of the network is used directly, limiting the maximum velocity between -1m/s and 1m/s. However, we also included the possibility for a ``simulated" attacker to boost the velocity of the high-performance controller by multiplying its output by a factor $\eta$ defined in the topic \texttt{/ext/v\_boost} depicted in Figure \ref{f:rover_ros2}.
The network was trained as the actor of a deep RL agent, specifically using Deep Deterministic Policy Gradients (DDPG)~\cite{lillicrap2015continuous}. All the hyperparameters were initialized with the typical values, as found in \url{https://github.com/ghliu/pytorch-ddpg}.
The actor was trained on a custom environment that simulates LiDAR-based corridor navigation with randomly-generated paths of maximum length of 20m.
The reward at each step $r$ was set to 
%\begin{equation}
    $r = D - 1 + E$,
%\end{equation}
%\noindent 
where $D$ is the forward distance toward the goal traveled by the rover
%traveled by the rover in the episode 
and $E$ is the term added at the end of the episode defined as follows: $E=-100$ if a collision occurs (which terminates the episode right away);  $E=300$ if the rover reaches the goal (i.e., end of the corridor) before the maximum number of steps in the episode (1000 steps, sampling time 100 ms); $E=0$ otherwise. The -1 term is useful in the initial training phase to induce the rover to move forward rather than staying still to avoid collisions. Please note that the reward is not optimized for an optimal behavior of the rover, since the main contribution of the paper is the architectural configuration to enhance safety.
The high-performance controller was seamlessly tested and integrated on the real rover with no need for fine-tuning. The camera is used to detect a specific QR code placed at the end of the corridor. When that QR code is detected, the rover stops and starts rotating on itself, signaling the completed course.

\subsection{Results}\label{ss:rover_results}
The experiments were performed in the corridors of our laboratory. The selected course is about 20 meters long with a right turn and a left turn similar to the one shown in Figure~\ref{f:rover_system}. The QR code is printed on an A4 sheet and placed so that it is visible given the hight of the camera.
All the data from the experiments have been recorded as \emph{rosbags} that capture the topics of interest.

\paragraph{Validation of the safety zones} The mechanism of the safety zones is the key concept that allows keeping the overall system safe, i.e., without collisions. The threshold of each zone must be carefully selected to account for the breaking space required for a complete stop.

\begin{figure}
    \centering
    \includegraphics[width=0.9\linewidth]{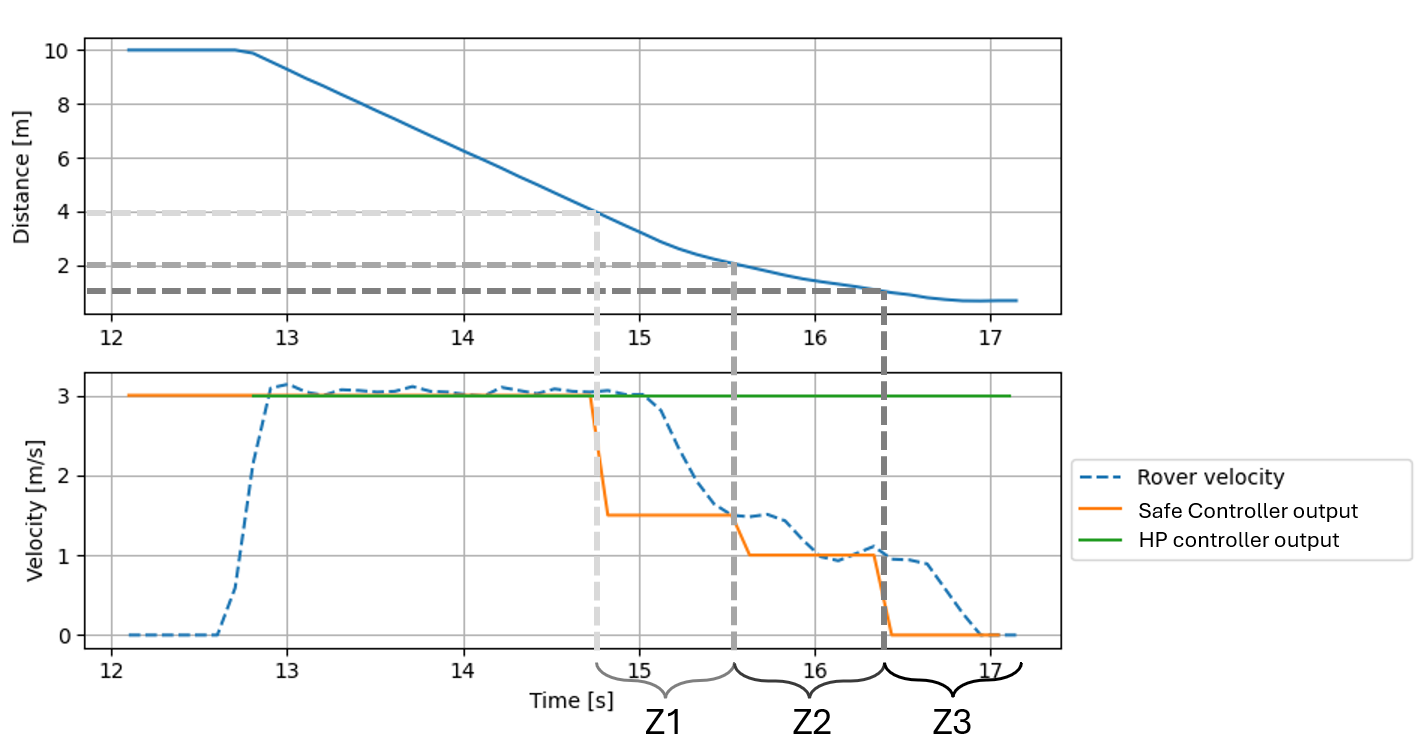}
    \caption{Validation of the safety zones. The graph above shows the distance to the front wall when the rover is controlled at the maximum velocity by the high-performance controller. The graph below illustrates how the requested velocity is regulated according to the safety zones. The rover measured velocity is also showed (dashed black) to visualize the breaking dynamics that clearly requires a certain margin to avoid collision.}
    \label{f:exp_rover_safety_zones}
\end{figure}

Hence, it is crucial that the safety zones keep the rover away from colliding with obstacles also in the worst case, i.e., when the rover is traveling at its maximum velocity, which is about 3 m/s.
The validation of the safety zones for the frontal distance was done by setting the high-performance controller to output a constant velocity close to the maximum velocity of the rover when facing a wall or an obstacle. 
If the safety zones are designed correctly, there should be no collision, since the progressively stringent safety zones will decrease the maximum velocity requested to the rover.
Figure~\ref{f:exp_rover_safety_zones} illustrates this process showing how a certain distance margin must be considered to take into account the breaking space
of the rover. This configuration is sufficient for our purposes. 

The same considerations hold for the lateral thresholds as well. However, for the scope of this paper, they were intentionally kept smaller, since the corridor is about two meters wide and more conservative safety zones would have triggered many false-positive safety stops. Also, considering that the high-performance controller's neural network has a \emph{tanh} activation, its output will always be between -1 and 1 m/s, which is slow enough for the system to be able to avoid collision with these safety zones.

\paragraph{Sample run}

\begin{figure}
    \centering
    \includegraphics[width=0.75\linewidth]{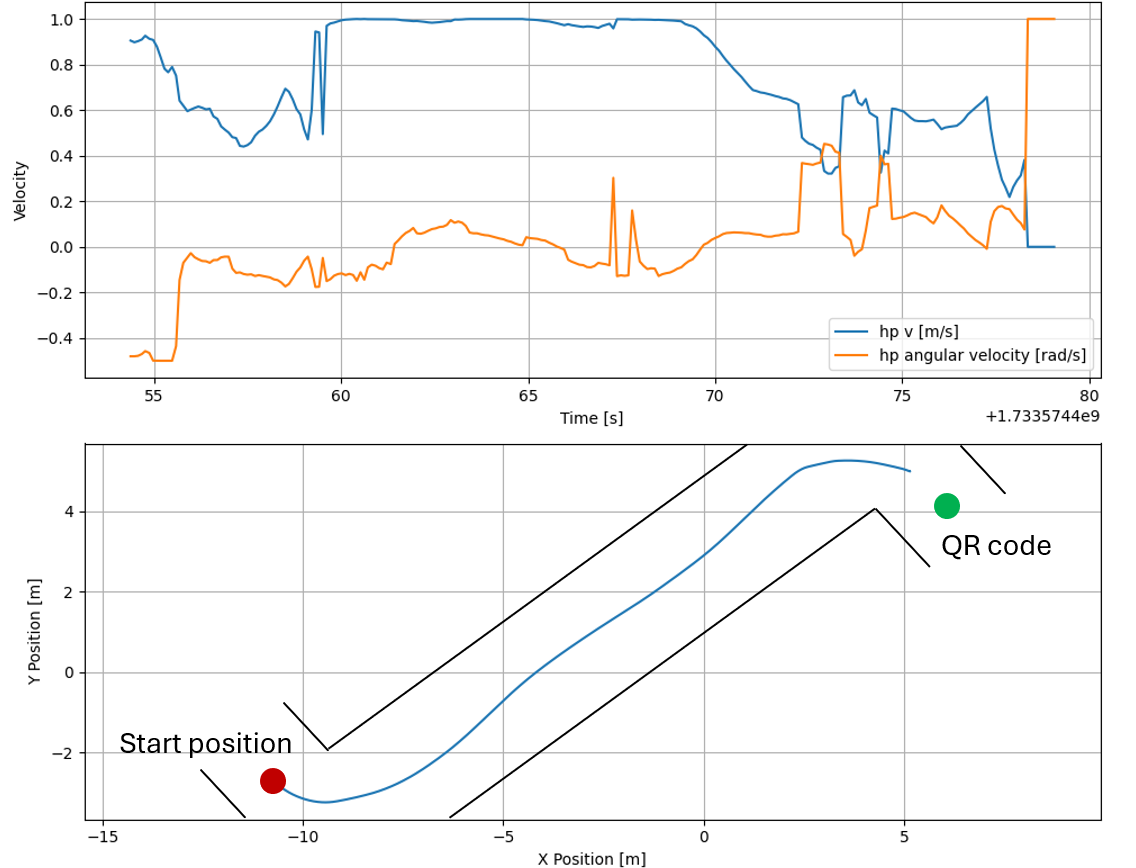}
    \caption{Velocity command from the high-performance controller (above) and position in the corridor (below) during a sample run with no safety stops triggered. The rover reaches the QR code without issues. When the QR code is detected, the rover is spun on itself at 1 rad/s.}
    \label{f:exp_rover_sample_run}
\end{figure}

Figure~\ref{f:exp_rover_sample_run} shows the output of the high-performance controller, which controls the velocity and the position of the rover in the corridor. In this run, no safety stop is triggered and the QR code is reached seamlessly. The safety zones are not visualized, as the velocity limitations are not affecting the high-performance controller: since the maximum velocity that the neural network can output with no external velocity boost is 1 m/s in standard condition, even when in Z2 the safe controller cannot limit the velocity unless a safety stop is triggered.

Since the system consistently presented a good behavior in normal conditions (including the avoidance of small obstacles placed close to the walls), we introduced two different dangerous situations to stress the systems in terms of safety and security: (i) sudden obstacles, and (ii) a simulated attack that modifies the output of the network.

\paragraph{Sudden obstacles}

This experiment is meant to validate the behavior of the rover in the presence of obstacles that might suddenly appear and could not be avoided by going around. Specifically, a cardboard box is thrown in front of the rover while navigating the corridor. 

\begin{figure}
    \centering
    \includegraphics[width=1.1\linewidth]{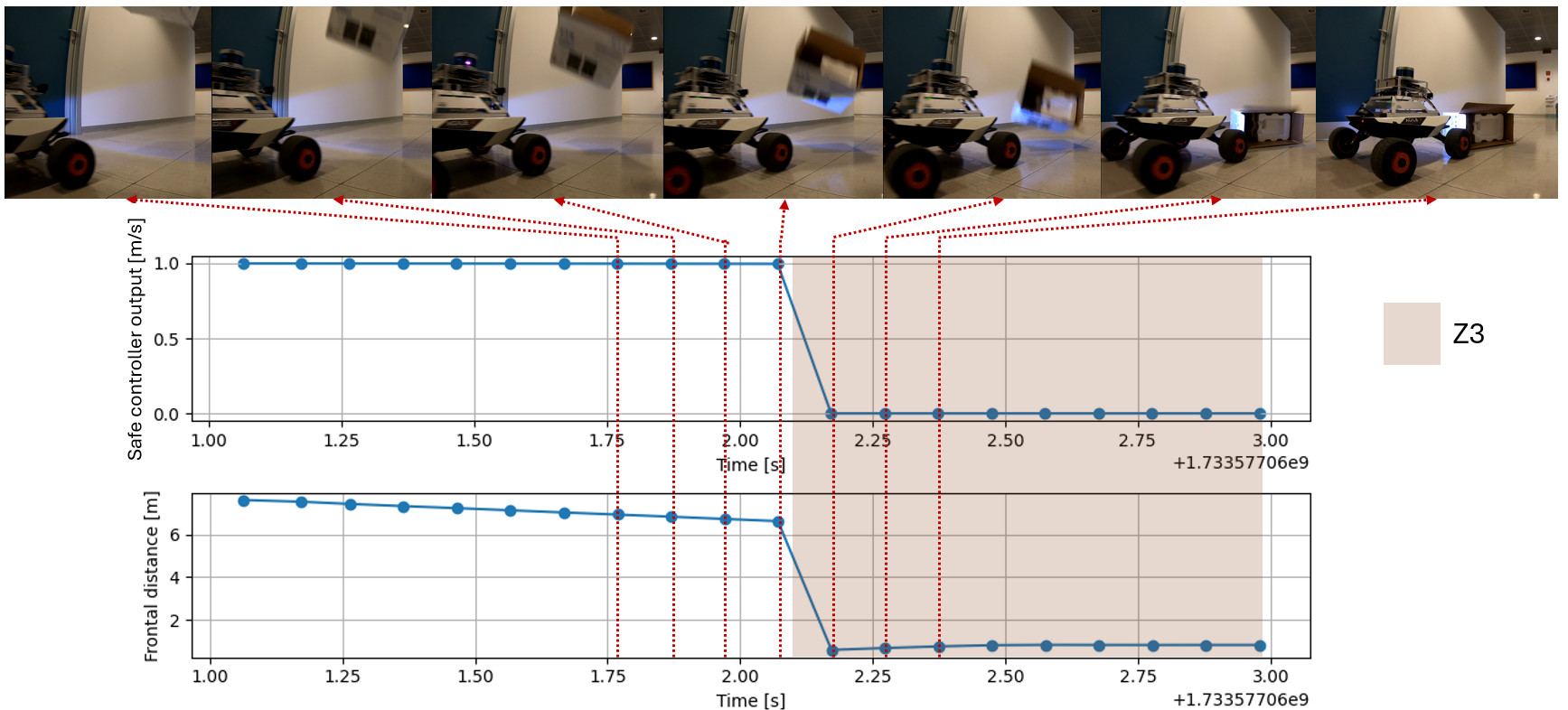}
    \caption{Velocity command (above) and frontal distance to the closest obstacle (below) during a run with a sudden obstacle appearance. The rover reacts within its 100 ms period to trigger a safety stop.}
    \label{f:exp_rover_obstacle}
\end{figure}

Figure~\ref{f:exp_rover_obstacle} shows the frontal distance extracted from the LiDAR and the corresponding actuation command. As soon as the obstacle is detected, a safety stop is triggered within the 100 ms LiDAR acquisition period. 

\paragraph{Simulated attack}

The \texttt{Safe\_VM} does not have internet access and cannot be interfered with. Conversely, the \texttt{Rich\_VM} can access internet for external visualization. This channel can be used to simulate a possible cyber-attack on the \texttt{Rich\_VM}. This test shows that the isolation property enforced by the hypervisor on the two domains and the safety zones mechanism prevent (or at least limit) unsafe behaviors from the high-performance controller.

Specifically, the high-performance controller is subscribed to a float32 topic that acts as a multiplicative factor for the output of the neural network. In this way, when an external user publishes a factor $\eta$ in the topic, the output of the neural network will no longer be limited between -1 and 1, but between $-\eta$ and $\eta$. Although this kind of attack might not be a proper cyber-attack, it is useful to illustrate how the safe controller takes care of unsafe velocity inputs from the high-performance controller. 

\begin{figure}
    \centering
    \includegraphics[width=\linewidth]{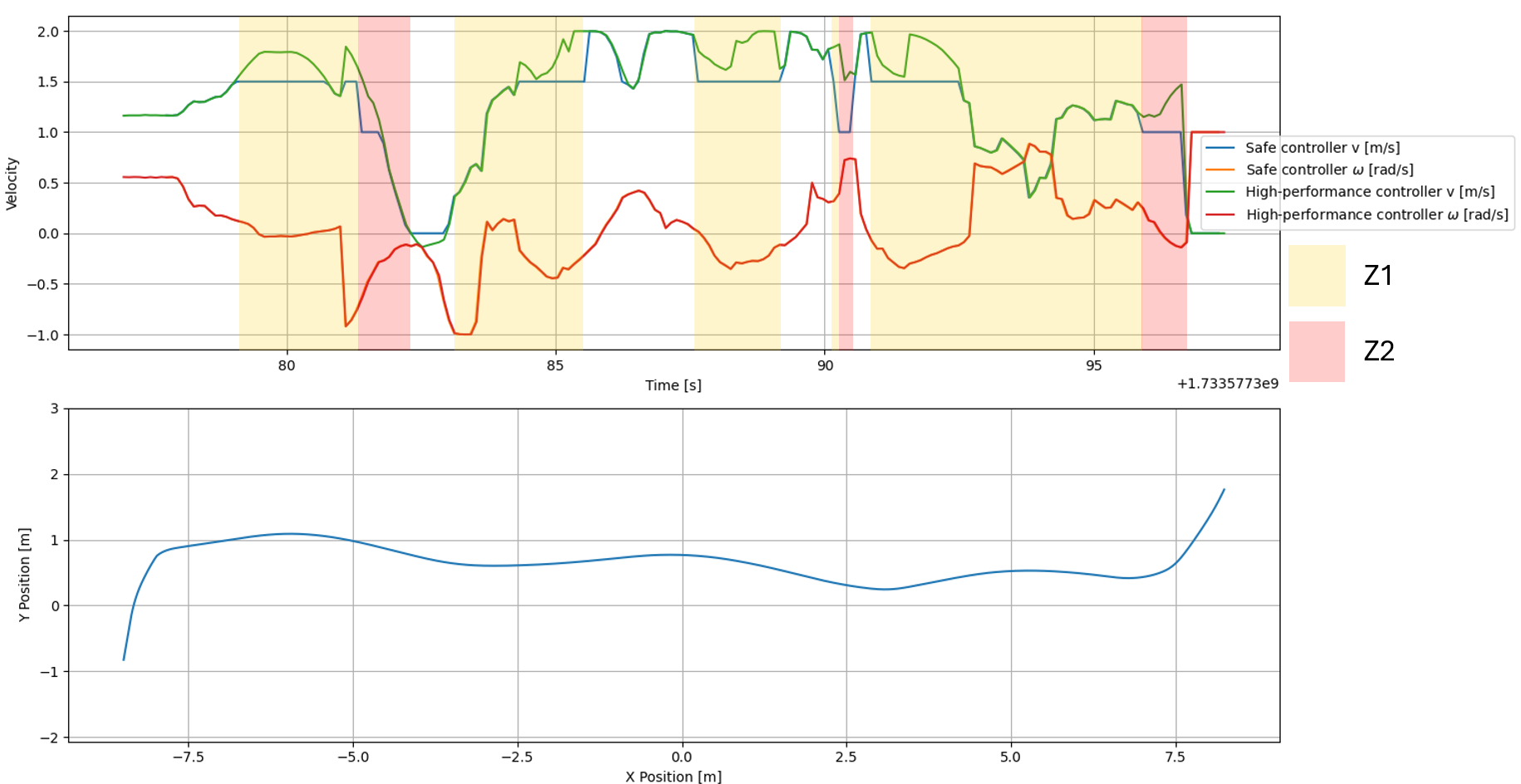}
    \caption{Velocity command from the high-performance controller and actual actuation value (above) and position in the corridor (below) during a run with $\eta=2$. The rover reaches the QR code, but there are a few situations where the safety zones limit the high-performance controller command, but never triggering a safety stop. }
    \label{f:exp_rover_attack}
\end{figure}

For this experiment, the rover is asked to complete the same course with $\eta = 2$. The results reported in Figure~\ref{f:exp_rover_attack} show that the rover is able to reach the QR code without safety stops. However, since now the maximum velocity from the high-performance controller is 2 m/s, the safety zones are entered and hence the actual velocity command is limited accordingly. The safety zones are triggered during the left and right curves of the corridor; also, on the straight part, the rover shows a much more ``wavy" behavior that triggers Z1 (and Z2 for a brief moment) because the walls were getting too close. This behavior of the neural network is likely arising because the agent has never traveled at such high velocities, while at low velocities the wave pattern is less visible.

\begin{figure}
    \centering
    \begin{subfigure}{0.455\columnwidth}
        \includegraphics[width=\columnwidth]{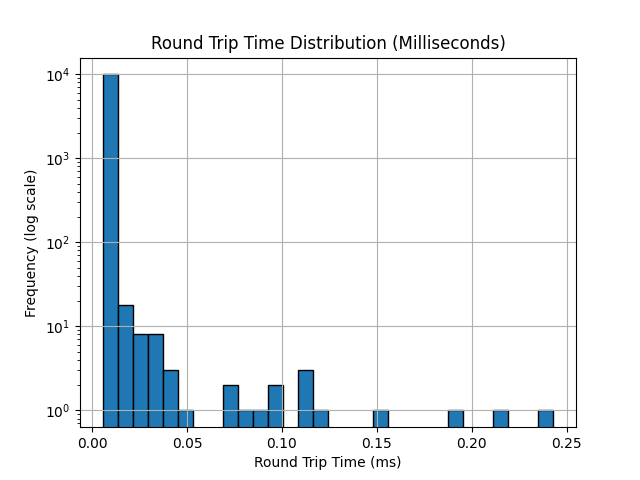}
    \caption{}
    \label{f:round_trip_comm_times}
    \end{subfigure}
    \begin{subfigure}{0.49\columnwidth}
    \includegraphics[width=\columnwidth]{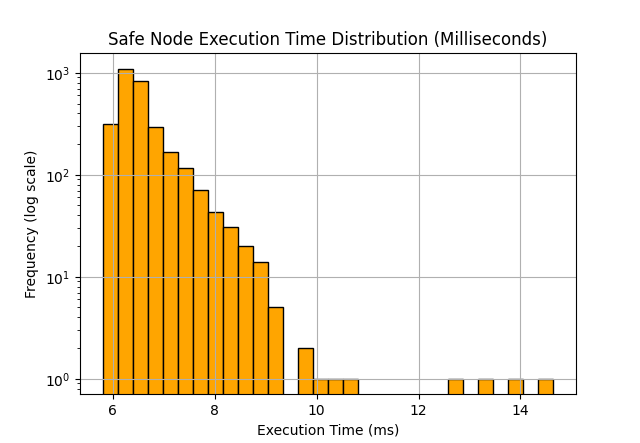}
    \caption{}
    \label{f:execution_time}
    \end{subfigure}
    \caption{(a) Round-trip communication times between the two domains for a 64-byte package (10.000 runs). (b) Distribution of the execution times of the \texttt{safe\_node} (3000 runs).}
    \label{f:temporal_analysis}
\end{figure}

\paragraph{Temporal analysis}
% TODO Niko
% 1) round-trip communication overhead
Figure \ref{f:round_trip_comm_times} shows the distribution of 10.000 measurements of the overhead introduced by the round-trip communication of a 64-byte package (which is more than enough for our purposes) between the safe and the rich domain and back. From the histogram it is clear that the communication between the two domains introduces a negligible overhead, with more than 99\% of the data passed in 10 microseconds or less.

Figure \ref{f:execution_time} shows the distribution of execution times for the \texttt{safe\_node} (3000 samples). This includes the time required to send LiDAR distances to the \texttt{Rich\_VM}, perform inference using the high-performance controller's neural network, and receive its output. From this distribution, we can infer that this implementation consistently meets the safe domain deadline of 100ms. 

% 2) compute safe node execution/response times
% Figure XXX shows the distribution of the execution times of the safe\_node. 

\section{Discussion, lessons learned, and conclusions}
\label{s:conclusions}

% DISCUSSION
% recap of the paper and contributions
This paper presented an architecture for CPSs that is designed to enhance the safety, security, and predictability levels of systems controlled with learning-based components.
Specifically, the main concept behind such an architecture is to separate the AI components from the safety-critical ones by using advanced virtualization technologies (provided by the CLARE-hypervisor). The AI components are always deemed untrustworthy, and a Simplex-like switching mechanism is designed to intervene whenever required to recover the safety and security levels, by excluding the learning-based components and their execution domain.

The paper also presented two fully functional implementations of the architecture: an inverted pendulum, used to highlight the main features of the architecture, and a rover obstacle avoidance application, representing a much more complex and complete use-case that showcases the main advantages of such an approach.

% limitations of the approach vs advantages
Despite the advantages discussed throughout the paper, this architecture clearly introduces non-negligible complexity into the hardware and software design. Specifically, the introduction of hypervisor technology requires specific know-how and dedicated configuration and programming tools; moreover, the use of different operating systems on the same platform seriously constrains the resources available to both domains.

\textcolor{black}{While for the pendulum the configuration is rather simple and device assignment did not cause any issue, the case of the rover was much more complex. In fact, each sensor had to be connected through a specific port (Ethernet for the LiDAR, USB for the motors) and, since these ports are also used by the rich domain for camera and internet access, these devices have to be partitioned between the two domains,
%This severely limits the amount of devices that can be attached to such ports,
complicating both the design and testing/debugging phases.}
\textcolor{black}{Moreover, for the pendulum, a single CPU core was enough to execute the safe domain tasks, allowing the use of three cores for the high-performance controller; for the rover, the safe domain was much more complex, requiring two cores to run ROS2 and all the assigned tasks on Linux. Device assignment required an iterative process to properly dimension the system.}
%--------------------------------------------
%GB: La frase seguente confonde: più potente ma con decreased performance??? 
%--------------------------------------------
%Typically, new-generation hardware is powerful enough to allow this increased complexity in spite of decreased performance. 
%--------------------------------------------

\textcolor{black}{Additional complexity is required to manage the inter-domain dedicated communication, which also introduces a certain latency. Such a latency is practically negligible in the case of the rover (for which the deadline is 100 ms), but may take up a significant fraction of the period for high-rate control systems such as the pendulum (4 ms). For this reasons, the periods of these tasks also required to be carefully engineered.}
\textcolor{black}{The use of ROS2 is another important addition in the case of the rover with respect to the case of the pendulum. ROS2 allows using pre-compiled packages to read the sensors and control the motion of the rover; on the other hand, it introduces non-negligible complexity, and, sometimes, unexpected behavior, probably due to the intricate code-base of the middleware. This was found to be especially complicated when dealing with the custom transport between the safe and rich domains, which required several iterations to work properly and with low latency. The case of the pendulum was simpler in this sense, while it required custom-designed FPGA logic to sample the encoders.}

Another point worth discussing is that controlling a system with two different, possibly alternating controllers requires specific precautions such as the introduction of a finite state machine that keeps the system in a safe state. 

\textcolor{black}{From a functional point of view, the safety monitors designed for the two cases are based on different concepts and are selected according to the definition of safety for the task at hand: the pendulum is monitored with a model-based technique built on a strong theoretical support, while the rover is monitored using LiDAR data that define a safety envelope to avoid crashes. It is worth noting how the Simplex architecture is adaptable to different safety monitoring methods, which might also include additional logic conditions, such as cyber-security warning systems that detect whether the rich domain is attacked.}
\textcolor{black}{All these downsides might slow down the design and development process, but the final tuned architecture significantly enhances the safety and security levels of autonomous CPSs, as demonstrated by our experiments. }

% FUTURE WORK
{\color{black}
Future work will investigate several different directions, including:
\begin{itemize}
    \item \textbf{The role of heterogeneous hardware for neural network acceleration}. FPGA is a solid alternative to the GPU for accelerating deep learning models. For both case studies, shallow networks were sufficient to showcase the proposed architecture, hence they were executed on the CPU without temporal issues. However, more complex visual tasks rely on deeper neural networks that require acceleration to work in real time. For this reason, future work will consider FPGA-accelerated DNNs in the implementation \cite{10.1145/3709026.3709107} \cite{11047633}.
    
    \item \textbf{Implementing vehicle-in-the-loop simulation.} Autonomous vehicles require extensive testing. However, it is often impossible to cover all the possible critical scenarios to validate the entire software stack. Vehicle-in-the-loop simulation tackles this problem by creating a digital twin that replicates the movements of the physical vehicle in a virtual world. The synthetic sensor data can be sent in real-time to the rich domain, which would act in the physical world based on simulated sensory inputs. %This feature will be explored in a future work.
    %MM: Ha senso l'ultima frase considerando che l'elenco puntato parte con "Future work will investigate"?
    
    \item \textbf{Evaluating the robustness of visual perception models.} Since DNNs are prone to adversarial attacks, the proposed architecture can be used to evaluate the robustness of the models to physical adversarial attacks. Also, detection algorithms can be used as triggers for the safety monitor.
    
    \item \textbf{Benchmarking the real-time and power properties of the system.} All the previous points (especially the vehicle-in-the-loop simulation) will require a careful analysis of the latencies introduced by the isolation. Including larger (accelerated) models will certainly bring more complexity into the design, which should always be backed by a comprehensive timing analysis. Furthermore, the power consumption overhead introduced by this approach will be investigated.

    \item \textbf{Adding novel recovery and security features.} The hypervisor can be used in conjunction with an anomaly detection system \cite{10.1145/3576914.3587551} deployed within the VMs to identify anomalies and restore the affected VM from a clean backup.

\end{itemize}
}



%% References


\section*{Acknowledgements}
This work has been partially supported by project SERICS (PE00000014) under the NRRP MUR program funded by the EU - NGEU and project RETICULATE of the PRIN 2022 program.

\bibliographystyle{elsarticle-num}
\bibliography{bibliography}

\begin{thebibliography}{10}
\expandafter\ifx\csname url\endcsname\relax
  \def\url#1{\texttt{#1}}\fi
\expandafter\ifx\csname urlprefix\endcsname\relax\def\urlprefix{URL }\fi
\expandafter\ifx\csname href\endcsname\relax
  \def\href#1#2{#2} \def\path#1{#1}\fi

\bibitem{Masana22}
M.~Masana, X.~Liu, B.~Twardowski, M.~Menta, A.~D. Bagdanov, J.~Van De~Weijer, Class-incremental learning: survey and performance evaluation on image classification, IEEE Transactions on Pattern Analysis and Machine Intelligence 45~(5) (2022) 5513--5533.

\bibitem{Zou23}
Z.~Zou, K.~Chen, Z.~Shi, Y.~Guo, J.~Ye, Object detection in 20 years: A survey, Proceedings of the IEEE 111~(3) (2023) 257--276.

\bibitem{Rybczak24}
M.~Rybczak, N.~Popowniak, A.~Lazarowska, A survey of machine learning approaches for mobile robot control, Robotics 13~(1) (2024) 12.

\bibitem{quinonero2022dataset}
J.~Qui{\~n}onero-Candela, M.~Sugiyama, A.~Schwaighofer, N.~D. Lawrence, Dataset shift in machine learning, Mit Press, 2022.

\bibitem{Sze14}
C.~Szegedy, W.~Zaremba, I.~Sutskever, J.~Bruna, D.~Erhan, I.~Goodfellow, R.~Fergus, Intriguing properties of neural networks, in: Proc. of the 2nd International Conference on Learning Representations (ICLR 2014), Banff, AB, Canada, April 14-16, 2014.

\bibitem{Big18}
B.~Biggio, F.~Roli, {Wild patterns: Ten years after the rise of adversarial machine learning}, Pattern Recognition 84 (2018) 317--331.

\bibitem{Kur17}
A.~Kurakin, I.~J. Goodfellow, S.~Bengio, Adversarial examples in the physical world, in: Proc. of the 5th Int. Conference on Learning Representation (ICLR), Toulon, France, April 2017.

\bibitem{Nesti_2022_WACV}
F.~Nesti, G.~Rossolini, S.~Nair, A.~Biondi, G.~Buttazzo, Evaluating the robustness of semantic segmentation for autonomous driving against real-world adversarial patch attacks, in: Proceedings of the IEEE/CVF Winter Conference on Applications of Computer Vision (WACV), 2022, pp. 2280--2289.

\bibitem{Cas20-spe}
D.~Casini, A.~Biondi, G.~Buttazzo, {Timing Isolation and Improved Scheduling of Deep Neural Networks for Real-Time Systems}, Software: Practice and Experience 50~(9) (2020) 1760--1777.

\bibitem{Cav17}
R.~Cavicchioli, N.~Capodieci, M.~Bertogna, {Memory Interference Characterization Between CPU Cores and Integrated GPUs in Mixed-Criticality Platforms}, in: Proc. of the 22nd IEEE International Conference on Emerging Technologies and Factory Automation (ETFA 2017), Limassol, Cyprus, September 12-15, 2017.

\bibitem{But22}
G.~Buttazzo, {Can We Trust AI-Powered Real-Time Embedded Systems?}, in: OpenAccess Series in Informatics (OASIcs), Vol. 98, Proc. of the HiPEAC Workshop on Next Generation Real-Time Embedded Systems (NG-RES 2022), Budapest, Hungary, June 22, 2022.

\bibitem{Clare}
{Accelerat Srl}, \href{https://accelerat.eu/clare}{{The CLARE Software Stack}}.
\newline\urlprefix\url{https://accelerat.eu/clare}

\bibitem{Mod18}
P.~Modica, A.~Biondi, G.~Buttazzo, A.~Patel, {Supporting Temporal and Spatial Isolation in a Hypervisor for ARM Multicore Platforms}, in: Proceedings of the 18th IEEE International Conference on Industrial Technology (ICIT 2018), Lyon, France, February 20-22, 2018.

\bibitem{Sha94}
L.~Sha, R.~Rajkumar, M.~Gagliardi, The simplex architecture: An approach to build evolving industrial computing systems, in: Proceedings of The ISSAT Int. Conference: Reliability and Quality in Design, Seattle, Washington, USA, March 16-18, 1994.

\bibitem{Bak09}
S.~Bak, D.~K. Chivukula, O.~Adekunle, M.~Sun, M.~Caccamo, L.~Sha, The system-level simplex architecture for improved real-time embedded system safety, in: Proceedings of the 15th IEEE Real-Time and Embedded Technology and Applications Symposium (RTAS’09), San Francisco, CA, USA, April 13-16, 2009.

\bibitem{Moh13}
S.~Mohan, S.~Bak, E.~Betti, H.~Yun, L.~Sha, M.~Caccamo, {S3A: Secure System Simplex Architecture for Enhanced Security and Robustness of Cyber-Physical Systems}, in: Proceedings of the ACM Int. Conference on High Confidence Networked Systems (HiCoNS), Philadelphia, PA, USA, April 9-11, 2013.

\bibitem{pereira2020challenges}
A.~Pereira, C.~Thomas, Challenges of machine learning applied to safety-critical cyber-physical systems, Machine Learning and Knowledge Extraction 2~(4) (2020) 579--602.

\bibitem{olowononi2020resilient}
F.~O. Olowononi, D.~B. Rawat, C.~Liu, Resilient machine learning for networked cyber physical systems: A survey for machine learning security to securing machine learning for cps, IEEE Communications Surveys \& Tutorials 23~(1) (2020) 524--552.

\bibitem{huang2020survey}
X.~Huang, D.~Kroening, W.~Ruan, J.~Sharp, Y.~Sun, E.~Thamo, M.~Wu, X.~Yi, A survey of safety and trustworthiness of deep neural networks: Verification, testing, adversarial attack and defence, and interpretability, Computer Science Review 37 (2020) 100270.

\bibitem{reke2020self}
M.~Reke, D.~Peter, J.~Schulte-Tigges, S.~Schiffer, A.~Ferrein, T.~Walter, D.~Matheis, A self-driving car architecture in ros2, in: 2020 International SAUPEC/RobMech/PRASA Conference, IEEE, 2020, pp. 1--6.

\bibitem{liu2017autopilot}
M.~Liu, J.~Niu, X.~Wang, An autopilot system based on ros distributed architecture and deep learning, in: 2017 IEEE 15th International Conference on Industrial Informatics (INDIN), IEEE, 2017, pp. 1229--1234.

\bibitem{gutierrez2018towards}
C.~S.~V. Guti{\'e}rrez, L.~U.~S. Juan, I.~Z. Ugarte, V.~M. Vilches, Towards a distributed and real-time framework for robots: Evaluation of ros 2.0 communications for real-time robotic applications, arXiv preprint arXiv:1809.02595 (2018).

\bibitem{o2019f1}
M.~O'Kelly, V.~Sukhil, H.~Abbas, J.~Harkins, C.~Kao, Y.~V. Pant, R.~Mangharam, D.~Agarwal, M.~Behl, P.~Burgio, et~al., F1/10: An open-source autonomous cyber-physical platform, arXiv preprint arXiv:1901.08567 (2019).

\bibitem{meier2015px4}
L.~Meier, D.~Honegger, M.~Pollefeys, Px4: A node-based multithreaded open source robotics framework for deeply embedded platforms, in: 2015 IEEE international conference on robotics and automation (ICRA), IEEE, 2015, pp. 6235--6240.

\bibitem{cittadini2023supporting}
E.~Cittadini, M.~Marinoni, A.~Biondi, G.~Cicero, G.~Buttazzo, Supporting ai-powered real-time cyber-physical systems on heterogeneous platforms via hypervisor technology, Real-Time Systems 59~(4) (2023) 609--635.

\bibitem{klein2018formally}
G.~Klein, J.~Andronick, M.~Fernandez, I.~Kuz, T.~Murray, G.~Heiser, Formally verified software in the real world, Communications of the ACM 61~(10) (2018) 68--77.

\bibitem{almeida2009safe}
J.~Almeida, M.~Prochazka, Safe and secure partitioning with pikeos: towards integrated modular avionics in space, DASIA 2009-DAta Systems in Aerospace 669 (2009) 27.

\bibitem{craveiro2009flexible}
J.~Craveiro, J.~Rufino, T.~Schoofs, J.~Windsor, Flexible operating system integration in partitioned aerospace systems, in: Actas do INForum-Simposio de Informatica, 2009, pp. 49--60.

\bibitem{perez2017handling}
H.~Perez, J.~J. Guti{\'e}rrez, Handling heterogeneous partitioned systems through arinc-653 and dds, Computer Standards \& Interfaces 50 (2017) 258--268.

\bibitem{farrukh2022flyos}
A.~Farrukh, R.~West, Flyos: Integrated modular avionics for autonomous multicopters, in: 2022 IEEE 28th Real-Time and Embedded Technology and Applications Symposium (RTAS), IEEE, 2022, pp. 68--81.

\bibitem{biondi2021sphere}
A.~Biondi, D.~Casini, G.~Cicero, N.~Borgioli, G.~Buttazzo, G.~Patti, L.~Leonardi, L.~L. Bello, M.~Solieri, P.~Burgio, et~al., Sphere: A multi-soc architecture for next-generation cyber-physical systems based on heterogeneous platforms, IEEE Access 9 (2021) 75446--75459.

\bibitem{Seto98}
D.~{Seto}, B.~{Krogh}, L.~{Sha}, A.~{Chutinan}, The simplex architecture for safe online control system upgrades, in: Proceedings of the 1998 American Control Conference. ACC (IEEE Cat. No.98CH36207), Vol.~6, 1998, pp. 3504--3508 vol.6.
\newblock \href {https://doi.org/10.1109/ACC.1998.703255} {\path{doi:10.1109/ACC.1998.703255}}.

\bibitem{Bak11}
S.~{Bak}, K.~{Manamcheri}, S.~{Mitra}, M.~{Caccamo}, Sandboxing controllers for cyber-physical systems, in: 2011 IEEE/ACM Second International Conference on Cyber-Physical Systems, 2011, pp. 3--12.
\newblock \href {https://doi.org/10.1109/ICCPS.2011.25} {\path{doi:10.1109/ICCPS.2011.25}}.

\bibitem{mohan2013s3a}
S.~Mohan, S.~Bak, E.~Betti, H.~Yun, L.~Sha, M.~Caccamo, S3a: Secure system simplex architecture for enhanced security and robustness of cyber-physical systems, in: Proceedings of the 2nd ACM international conference on High confidence networked systems, 2013, pp. 65--74.

\bibitem{Vivekanandan16}
K.~Vivekanandan, et~al., A simplex architecture for intelligent and safe unmanned aerial vehicles, in: IEEE International Conference on Embedded and Real-Time Computing Systems and Applications, 2016.

\bibitem{Bio20-esl}
A.~Biondi, F.~Nesti, G.~Cicero, D.~Casini, G.~Buttazzo, {A Safe, Secure, and Predictable Software Architecture for Deep Learning in Safety-Critical Systems}, IEEE Embedded Systems Letters 12~(3) (2020) 78--82.

\bibitem{desai2019soter}
A.~Desai, S.~Ghosh, S.~A. Seshia, N.~Shankar, A.~Tiwari, Soter: a runtime assurance framework for programming safe robotics systems, in: 2019 49th Annual IEEE/IFIP International Conference on Dependable Systems and Networks (DSN), IEEE, 2019, pp. 138--150.

\bibitem{phan2020neural}
D.~T. Phan, R.~Grosu, N.~Jansen, N.~Paoletti, S.~A. Smolka, S.~D. Stoller, Neural simplex architecture, in: NASA Formal Methods: 12th International Symposium, NFM 2020, Moffett Field, CA, USA, May 11--15, 2020, Proceedings 12, Springer, 2020, pp. 97--114.

\bibitem{Zubov65}
V.~I. Zubov, Methods of a. m. lyapunov and their application, The American Mathematical Monthly (December 1965).

\bibitem{Seto99}
D.~Seto, L.~Sha, A case study on analytical analysis of the inverted pendulum real-time control system, 1999.

\bibitem{Ferreira97}
E.~Ferreira, B.~Krogh, Using neural networks to estimate regions of stability, in: Proceedings of 1997 American Control Conference, Vol.~3, 1997, pp. 1989 -- 1993.

\bibitem{Ferreira99}
E.~Ferreira, B.~Krogh, Switching controllers based on neural network estimates of stability regions and controller performance, 1999.
\newblock \href {https://doi.org/10.1007/3-540-64358-3_36} {\path{doi:10.1007/3-540-64358-3_36}}.

\bibitem{Bak14}
S.~{Bak}, T.~T. {Johnson}, M.~{Caccamo}, L.~{Sha}, Real-time reachability for verified simplex design, in: 2014 IEEE Real-Time Systems Symposium, 2014, pp. 138--148.
\newblock \href {https://doi.org/10.1109/RTSS.2014.21} {\path{doi:10.1109/RTSS.2014.21}}.

\bibitem{abdar2021review}
M.~Abdar, F.~Pourpanah, S.~Hussain, D.~Rezazadegan, L.~Liu, M.~Ghavamzadeh, P.~Fieguth, X.~Cao, A.~Khosravi, U.~R. Acharya, et~al., A review of uncertainty quantification in deep learning: Techniques, applications and challenges, Information fusion 76 (2021) 243--297.

\bibitem{yang2024generalized}
J.~Yang, K.~Zhou, Y.~Li, Z.~Liu, Generalized out-of-distribution detection: A survey, International Journal of Computer Vision (2024) 1--28.

\bibitem{rossolini2023defending}
G.~Rossolini, F.~Nesti, F.~Brau, A.~Biondi, G.~Buttazzo, Defending from physically-realizable adversarial attacks through internal over-activation analysis, in: Proceedings of the AAAI Conference on Artificial Intelligence, Vol.~37, 2023, pp. 15064--15072.

\bibitem{zhu2015optimal}
F.~Zhu, P.~J. Antsaklis, Optimal control of hybrid switched systems: A brief survey, Discrete Event Dynamic Systems 25 (2015) 345--364.

\bibitem{Quanser}
Quanser, \href{https://www.quanser.com/products/rotary-inverted-pendulum/}{Quanser rotary inverted pendulum}.
\newline\urlprefix\url{https://www.quanser.com/products/rotary-inverted-pendulum/}

\bibitem{cazzolato2011dynamics}
B.~S. Cazzolato, Z.~Prime, On the dynamics of the furuta pendulum, Journal of Control Science and Engineering 2011~(1) (2011) 528341.

\bibitem{astrom}
K.~Åström, K.~Furuta, \href{http://www.sciencedirect.com/science/article/pii/S0005109899001405}{Swinging up a pendulum by energy control}, Automatica 36~(2) (2000) 287 -- 295.
\newblock \href {https://doi.org/https://doi.org/10.1016/S0005-1098(99)00140-5} {\path{doi:https://doi.org/10.1016/S0005-1098(99)00140-5}}.
\newline\urlprefix\url{http://www.sciencedirect.com/science/article/pii/S0005109899001405}

\bibitem{us_trm}
Xilinx, \href{https://docs.amd.com/r/en-US/ug1085-zynq-ultrascale-trm/Zynq-UltraScale-Device-Technical-Reference-Manual}{Zynq ultrascale+ device technical reference manual-ug1085}.
\newline\urlprefix\url{https://docs.amd.com/r/en-US/ug1085-zynq-ultrascale-trm/Zynq-UltraScale-Device-Technical-Reference-Manual}

\bibitem{Erika}
{Evidence Srl}, \href{https://www.erika-enterprise.com/}{{Erika Enterprise RTOS}}.
\newline\urlprefix\url{https://www.erika-enterprise.com/}

\bibitem{auger2012tutorial}
A.~Auger, N.~Hansen, Tutorial cma-es: evolution strategies and covariance matrix adaptation, in: Proceedings of the 14th annual conference companion on Genetic and evolutionary computation, 2012, pp. 827--848.

\bibitem{brockman2016gym}
G.~{Brockman}, V.~{Cheung}, L.~{Pettersson}, J.~{Schneider}, J.~{Schulman}, J.~{Tang}, W.~{Zaremba}, {OpenAI Gym}, arXiv e-prints (2016) arXiv:1606.01540\href {http://arxiv.org/abs/1606.01540} {\path{arXiv:1606.01540}}, \href {https://doi.org/10.48550/arXiv.1606.01540} {\path{doi:10.48550/arXiv.1606.01540}}.

\bibitem{AgileX}
AgileX, \href{https://global.agilex.ai/products/scout-mini}{Agilex scout mini}.
\newline\urlprefix\url{https://global.agilex.ai/products/scout-mini}

\bibitem{lizano2024comparison}
S.~A. Lizano, T.~Westerlund, Comparison of edge computing platforms for hardware acceleration of ai: Kria kv260, jetson nano and rtx 3060, J. Edge Comput. 15~(3) (2024) 123--135.

\bibitem{ros2}
S.~Macenski, T.~Foote, B.~Gerkey, C.~Lalancette, W.~Woodall, \href{https://www.science.org/doi/abs/10.1126/scirobotics.abm6074}{Robot operating system 2: Design, architecture, and uses in the wild}, Science Robotics 7~(66) (2022) eabm6074.
\newblock \href {https://doi.org/10.1126/scirobotics.abm6074} {\path{doi:10.1126/scirobotics.abm6074}}.
\newline\urlprefix\url{https://www.science.org/doi/abs/10.1126/scirobotics.abm6074}

\bibitem{lillicrap2015continuous}
T.~Lillicrap, Continuous control with deep reinforcement learning, arXiv preprint arXiv:1509.02971 (2015).

\bibitem{10.1145/3709026.3709107}
Z.~Bao, H.~Li, W.~Zhang, \href{https://doi.org/10.1145/3709026.3709107}{A vision transformer inference accelerator for kr260}, in: Proceedings of the 2024 8th International Conference on Computer Science and Artificial Intelligence, CSAI '24, Association for Computing Machinery, New York, NY, USA, 2025, p. 245–251.
\newblock \href {https://doi.org/10.1145/3709026.3709107} {\path{doi:10.1145/3709026.3709107}}.
\newline\urlprefix\url{https://doi.org/10.1145/3709026.3709107}

\bibitem{11047633}
Y.~Chen, L.~Yu, Fpga-based mcs-yolov5 lightweight ship detection modeltection on kr260, in: 2025 5th International Conference on Artificial Intelligence and Industrial Technology Applications (AIITA), 2025, pp. 1764--1768.
\newblock \href {https://doi.org/10.1109/AIITA65135.2025.11047633} {\path{doi:10.1109/AIITA65135.2025.11047633}}.

\bibitem{10.1145/3576914.3587551}
N.~Borgioli, L.~Thi Xuan~Phan, F.~Aromolo, A.~Biondi, G.~Buttazzo, \href{https://doi.org/10.1145/3576914.3587551}{Real-time packet-based intrusion detection on edge devices}, in: Proceedings of Cyber-Physical Systems and Internet of Things Week 2023, CPS-IoT Week '23, Association for Computing Machinery, New York, NY, USA, 2023, p. 234–240.
\newblock \href {https://doi.org/10.1145/3576914.3587551} {\path{doi:10.1145/3576914.3587551}}.
\newline\urlprefix\url{https://doi.org/10.1145/3576914.3587551}

\end{thebibliography}






% \end{thebibliography}

\end{document}